\documentclass[superscriptaddress,amsmath,amssymb,prd,preprintnumbers,showpacs,twocolumn,floatfix]{revtex4-2}
\usepackage{graphicx}
\usepackage[T1]{fontenc}
\usepackage{amssymb,amsmath,bm,natbib}
\usepackage{color}
\usepackage{slashed}
\usepackage{graphics}
\usepackage{graphicx}
\usepackage[utf8]{inputenc}
\usepackage[caption=false]{subfig}
\usepackage{hyperref}
\usepackage{url}
\usepackage{dsfont}
\usepackage{float}
\usepackage{cancel}
\usepackage{units}
\usepackage{blindtext}
\usepackage[utf8]{inputenc}
\usepackage{upgreek}
\usepackage{booktabs}
\usepackage[dvipsnames,table,xcdraw]{xcolor}
\usepackage{enumerate}
\usepackage{mathtools}
\usepackage{soul,color}
\usepackage[normalem]{ulem}
\usepackage{dblfloatfix}
\usepackage{placeins}
\usepackage[vcentermath]{youngtab}

\def\ring#1{{\mathaccent'27 #1}}

\newcommand{\orcid}[1]{\href{https://orcid.org/#1}{\includegraphics[width=10pt]{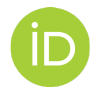}}}

\begin{document}

\title{New constraints on modified gravity with
 dimension-six operators\\ from gravitational waves}

\author{Justo L\'{o}pez-Sarri\'{o}n\orcid{0000-0002-2868-194X}}
\email{justo.lopezsarrion@unizar.es}
\affiliation{CAPA $\&$ Departamento de Física Teórica, Universidad de Zaragoza, Zaragoza 50009, Spain}
\author{Carlos M. Reyes\orcid{0000-0001-5140-6658}}
\email{creyes@ubiobio.cl}
\affiliation{Centro de Ciencias Exactas, Facultad de Ciencias, Universidad del B\'{i}o-B\'{i}o,
Chill\'{a}n, Casilla 447, Chile}

\author{C\'esar Riquelme\orcid{0000-0003-0837-3891}}
\email{ceriquelme@udec.cl}
\affiliation{Facultad de Ingenier\'{i}a, Universidad San Sebasti\'{a}n, Lientur 1457, Concepci\'{o}n, Chile}

\author{Marco Schreck\orcid{0000-0001-6585-4144}}
\email{marco.schreck@ufma.br}
\affiliation{Coordena\c{c}\~{a}o do Curso de F\'{i}sica -- Bacharelado, Universidade Federal do Maranh\~ao, Campus Universit\'{a}rio do Bacanga, S\~ao Lu\'is (MA), 65080-805, Brazil}
\affiliation{Programa de P\'{o}s-graduaç\~{a}o em F\'{i}sica, Universidade Federal do Maranh\~{a}o, Campus Universit\'{a}rio do Bacanga, S\~ao Lu\'is (MA), 65080-805, Brazil}

\begin{abstract}
The present article deals with gravitational-wave propagation affected by
higher-derivative terms that break spacetime symmetries. Two dimension-6
contributions of the gravitational Standard-Model Extension pose our starting
point. A linearization of the action implies a wave equation that contains additional terms
with spacetime-constant background fields. The modified dispersion relations
in covariant form for gravitational waves are derived, where we distinguish
between nonbirefringent and birefringent sectors. A classification of the
coefficients in terms of sets with index structures resembling those of
the electromagnetic fields has proven to be valuable. We constrain
the nonbirefringent coefficients based on the measured arrival time difference
between the gravitational wave and photons from the events GW170817 and GRB 170817A,
respectively. Bounds on the birefringent coefficients result from the
absence of a perceivable separation of the two modes in the event GW150914. These
findings quantify the extent to which standard linearized gravity
is valid based on the modifications considered.
\end{abstract}
\maketitle
\section{Introduction}
\label{sec:introduction}
After Einstein had found the field equations of gravity in their final form still used
today, he immediately thought about whether his theory exhibits wave-like
solutions~\cite{Einstein:1916cc,Einstein:1918btx}. For several decades, this question remained
unanswered in a satisfactory manner and was only settled in the 1950s. In fact, the
field equations of General Relativity (GR), when linearized properly around the
Minkowski metric, have solutions in the form of propagating plane waves. Yet,
they are of a different nature than the wave solutions in Maxwell electrodynamics. Instead
of describing periodic modulations of vector-valued electric and magnetic fields
propagating through spacetime, gravitational waves correspond to periodic
distortions of spacetime geometry itself.
 In analogy to electromagnetic waves, gravitational waves
have two physical polarizations known as the $+$ and $\times$ modes. They propagate
with the speed of gravity, which corresponds to the speed of light~\cite{Moore:2013}.

After it became accepted that linearized gravity possesses wave solutions, the
question arose as to whether and how they could possibly be detected experimentally.
First of all, gravitational waves are produced by a source subject to periodic
changes that are not spherically symmetric. In fact, measuring the rotation period
of the binary pulsar B1913+16 in the 1970s showed that it loses energy
by emitting gravitational waves~\cite{Weisberg:2004hi,Weisberg:2016jye}, precisely
in accordance with the linearized-gravity description. Therefore, this finding was
interpreted as an indirect experimental confirmation of the existence of gravitational waves.

Direct detection has proven to be extremely challenging. Potential sources
should provide signals strong enough that they have a chance of being
measured on Earth. The wave amplitude, known as the strain, decreases with the inverse of the
distance from the source, as does the field amplitude of electromagnetic waves. However, while detectors for electromagnetic waves measure intensity, which is proportional to the wave amplitude squared, gravitational-wave detectors directly measure the strain.
Suitable sources include binaries of neutron stars and
black holes, as well as mergers of the latter.

Interferometric approaches have proven to be the preferred direct-detection
methods since they allow us to achieve the incredible precision necessary.
In 2016, LIGO announced the first event~\cite{LIGOScientific:2016aoc} that
humanity had ever observed at that point. Since then, many additional
signals have complemented the first observation made, whereby the catalog
GWTC-3~\cite{KAGRA:2021vkt} is the most recent compilation of the LIGO
Scientific, Virgo, and KAGRA Collaborations.

Gravitational waves present exquisite events for testing gravity in the regime of
weak gravitational fields. Signals of physics beyond standard linearized
gravity include, but are not restricted to, propagation velocities different
from the speed of light, polarization-dependent propagation, anisotropies,
additional modes, etc. There are various possibilities for how to search for
deviations from standard linearized gravity based on Einstein's relativity.
One can work with specific models beyond GR, propose modifications
at the level of the dispersion relation, or choose a comprehensive test framework
such as the gravitational Standard-Model Extension~(SME)~\cite{Kostelecky:2003fs,Kostelecky:2020hbb}.
We will follow the latter approach.

The gravitational SME alters the usual Einstein-Hilbert (EH)
action by incorporating all terms invariant under general coordinate
transformations but violating spacetime symmetries. Symmetry violations
are parametrized by background fields that are composed of controlling coefficients. These are either present in the theory without knowing anything about their physical origin~\cite{Bluhm:2014oua,Bluhm:2016dzm,Bluhm:2017pje,
Bluhm:2019ato,Bluhm:2021lzf,Reyes:2021cpx,Reyes:2022mvm,Bluhm:2023kph} or they arise via a dynamical mechanism, such as in bumblebee models~\cite{Bluhm:2004ep,Bluhm:2014oua,Bluhm:2017pje,Bluhm:2023kph}. In the latter, a vector-valued vacuum expectation value $b_{\mu}=\langle B_{\mu}\rangle$ of a dynamical field $B_{\mu}$ emerges in a suitably chosen potential. Hybrids of both approaches~\cite{Bluhm:2015dna,Schreck:2024hky} have been proposed, too.

For explicit spacetime symmetry breaking, dynamics is known to exhibit tensions
with the second Bianchi identities of pseudo-Riemannian geometry~\cite{Kostelecky:2003fs,Kostelecky:2020hbb}.
However, it is still an interesting possibility that can generically be valid in a
beyond-Riemannian setting, such as Finsler geometry~\cite{Finsler:1918,Shen:2000}.
The SME community has been studying the latter possibility for several years~\cite{Kostelecky:2011qz,AlanKostelecky:2012yjr,
Russell:2015gwa,Schreck:2015seb,Edwards:2018lsn,Davis:2025die},
where relativistic point-particle Lagrangians~\cite{Kostelecky:2010hs,Colladay:2012rv,
Schreck:2014ama,Reis:2017ayl,Schreck:2019mmr,Reis:2021ban,Reis:2026rch}
have served as a foundation for developing these ideas. Additionally, several methods have
emerged to treat explicit symmetry violations consistently~\cite{Bluhm:2019ato,Reyes:2024ywe,Reyes:2024hqi,Riquelme:2026ztl}.

The minimal gravitational SME includes all field operators of mass dimensions $d\leq 4$,
whereas the nonminimal SME also permits operators of $d>5$. Laboratory
tests of short-range gravity have constrained certain nonrelativistic
combinations of $d=6$ operators~\cite{Long:2014swa,Shao:2015gua,Shao:2016cjk}.
The absence of gravitational Cherenkov radiation, i.e., the energy loss of
matter particles or even photons via the radiation of gravitational waves,
has put strict
limits on operators of $d=4,6,8$~\cite{Caves:1980jn,Moore:2001bv,
Elliott:2005va,Kimura:2011qn,DeLaurentis:2012fh,Kostelecky:2015dpa,Kiyota:2015dla}.
The data tables~\cite{Kostelecky:2008ts} provide a yearly-updated compilation
of bounds on SME coefficients. We note in passing that formal aspects in quantum field theories in Minkowski spacetime with nonminimal
spacetime symmetry violation, in particular, unitarity and stability, have been investigated, as well~\cite{Lopez-Sarrion:2023nux,Lopez-Sarrion:2022czo,Ferreira:2020wde,Reyes:2010pv}. Studies along these lines within a linearized-gravity regime could be envisioned.

Among the first papers proposing tests of spacetime symmetry violations via gravitational waves are Refs.~\cite{Mirshekari:2011yq,Hansen:2014ewa}. The former introduces a modified dispersion relation for gravitational waves, which includes a mass and an additional frequency dependence, such that Lorentz invariance is violated. Certain quantum-gravity models motivate modifications of this kind. The researchers then studied how gravitational-wave propagation is affected and inferred sensitivities for these alterations that future detectors might potentially achieve.

The second paper rests upon Einstein-\ae{}ther theory and khonometric gravity. The authors were interested in constraining the parameters of these theories via data on gravitational waves emitted by neutron star mergers in the late-inspiral phase. Although second-generation detectors do not imply better constraints than data on binary pulsars, third-generation detectors are capable of doing so. The paper also comments on multimessenger signals, e.g., the simultaneous emission of photons or neutrinos and gravitational waves by a single source. The potential of such events in constraining exotic physics is emphasized, which will also prove vital for our endeavors.

The phenomenological paper~\cite{Kostelecky:2016kfm} marks the advent of searches for signals of spacetime symmetry violation in gravitational waves within the SME. The authors determined generic dispersion relations for diffeomorphism-invariant and Lorentz-violating modifications of the linearized action. They then calculated the first constraints on $d=5,6$ operators from gravitational waves. Several theoretical papers followed afterward. One important initial goal was to classify all possible coordinate-invariant modifications of linearized gravity according to their transformation properties under diffeomorphisms and Lorentz transformations~\cite{Kostelecky:2017zob}. Mewes exhaustively studied the impact of Lorentz violation on polarization, waveforms, and the strain signal~\cite{Mewes:2019yyq,Mewes:2019dhj}. At around the same time, Xu was interested in finding out what the $s_{\mu\nu}$ coefficients do to polarization and the motion of test bodies hit by a modified gravitational wave~\cite{Xu:2019fyt}. In a related manner, the article~\cite{Schreck:2016qiz} reports on the impact of $a_{\mu}$ Lorentz violation in the matter sector on gravitational-wave detection.

These efforts continued in Refs.~\cite{Xu:2019gua,Xu:2020zxs,Xu:2021dcw}, whose authors explored how $s_{\mu\nu}$ affects spinning spherical and elliptical neutron stars, the quadrupole moment formula, and the emission of gravitational waves. Reference~\cite{Bailey:2023lzy} is dedicated to solving the gravitational-wave equation with matter and charge sources for $s_{\mu\nu}$. The article~\cite{Hou:2024xbv} delves into the theory of gravitational-wave polarization and detector response functions, which are deemed valuable for constraining the $d=4$ $s$-type coefficients introduced in Ref.~\cite{Kostelecky:2017zob}.

After gaining more control over the theory, the time was ripe for broad searches for spacetime symmetry violations in gravitational-wave data. The isotropy of gravitational-wave propagation was tested in Refs.~\cite{Liu:2020slm,Ray:2023sbr}, which resulted in limits on $s_{\mu\nu}$. Sensitivities for $s_{\mu\nu}$ as well as matter-gravity couplings from space-based detectors were estimated in Ref.~\cite{Qin:2023xvz}. The gravitational-wave equation modified by $s_{\mu\nu}$ was solved explicitly in Ref.~\cite{AouladLafkih:2025stw}, which allowed the researchers to constrain certain subsets of these coefficients. In their recent work~\cite{AraujoFilho:2026zyt}, the authors considered isotropic $d=4$ Lorentz violation with the intent of understanding their impact on gravitational-wave generation. These coefficients were bounded via the time delay between photons and gravitational waves from the multimessenger event GW170817/GRB 170817A, referred to in more detail later.

Operators of mass dimensions different from $d=4$ have also been taken into account in gravitational-wave studies. Coefficients of operators of mass dimensions 5 and 6 were bounded in Refs.~\cite{Shao:2020shv,ONeal-Ault:2021uwu,Wang:2021ctl,Niu:2022yhr,Haegel:2022ymk,Gong:2023ffb}. Reference~\cite{Wang:2025fhw} considers operators of $d=2\dots 6$, where studies of dephasing in modified waveforms gave rise to explicit constraints for $d=2,3$. Propagation effects altered by isotropic configurations of $d=4,5$ operators were investigated in Ref.~\cite{AraujoFilho:2026vcf}.

The authors of Ref.~\cite{Cao:2024lvd} proposed employing multimessenger data from binary white dwarfs. They estimated the sensitivity of the propagation speed difference between photons and gravitational waves, the graviton mass, and the $s_{00}$ coefficient. A similar project was carried out in Ref.~\cite{Rao:2024wde} via simulations of neutron star mergers with the events to be detected by Advanced LIGO and Einstein Telescope.

Bumblebee-type models have also been in the spotlight within gravitational-wave physics. First of all, a correction to the standard quadrupole formula for a purely spacelike~$b_{\mu}$ was obtained in Ref.~\cite{Amarilo:2018zqg}. Moreover, the authors of Ref.~\cite{Liang:2022hxd} investigated the polarization content in the bumblebee model, as well as for the minimal $u$ and $s_{\mu\nu}$ coefficients arising dynamically. The impact of the bumblebee field on gravitational-wave production was again examined in Ref.~\cite{Amarilo:2023wpn}, which resulted in a constraint on the characteristic dimensionless parameter~$\xi b^2$. In addition, there has been interest in the bumblebee model in cosmology, where multimessenger signals allowed for setting limits on $\xi b^2$~\cite{Lai:2025nyo}. In Ref.~\cite{Khodadi:2025wuw}, the sensitivity on $\xi b^2$ for a variety of (future) gravitational-wave detectors was estimated. Finally, there are papers on the propagation of test particles in black-hole solutions of the gravitational bumblebee model, which imply waveforms characteristic of such models to be tested in experiments~\cite{Shi:2026zxx}.

A significant number of papers have been written on searches for Lorentz violation beyond the SME. Some researchers constrain the free parameters of the modified gravitational-wave dispersion relation proposed in Refs.~\cite{Mirshekari:2011yq,Hansen:2014ewa}, which are the mass, the dimensionful prefactor of the frequency-dependent modification, and the dimensionless power of the frequency.
In the paper~\cite{LIGOScientific:2017bnn}, the previous parameters were bounded via a single event. In the forthcoming years, Advanced LIGO and Advanced Virgo collected a sufficient number of events to allow for an extended study~\cite{LIGOScientific:2019fpa,Samajdar:2019ptt,LIGOScientific:2020tif,LIGOScientific:2021sio,Haegel:2021tvb,Baka:2025drk}, which has led to refined constraints.

The velocity difference between gravitational waves and photons in a certain multimessenger signal was constrained in Ref.~\cite{Ellis:2016rrr}. The propagation of gravitational waves in a cosmological background was also studied in a setting known as spatially covariant gravity, which extends the action of GR at the level of the $(3+1)$ decomposition~\cite{Gao:2019liu,Zhu:2022uoq}. Modifications of gravitational-wave propagation in Ho\v{r}ava-Lifshitz gravity coupled to electromagnetism were bounded via multimessenger signals~\cite{Zhang:2020bzg}. Alternatively, some researchers started from a modified time evolution equation of gravitational-wave strain, which can be translated into an isotropic dispersion relation with higher-order frequency dependence~\cite{Gong:2021jgg,Zhu:2023rrx}. Interestingly, gravitational waves were studied in ghost-free massive gravity, which, beyond dispersion and anisotropies, exhibit additional peculiar effects such as pentarefringence~\cite{Kostelecky:2021xhb}. Research on higher-derivative Einstein-\ae{}ther theory and modified waveforms has also been carried out~\cite{Choudhury:2025qsh}.

The main goal of the current paper is to extend the previous analyses of
$d=6$ operators in the SME gravity sector. The alterations of the EH action
are of the symbolic forms $k_RRR$ and $k_D\lbrace \nabla,\nabla \rbrace R$ with the Riemann curvature tensor $R$, the covariant derivative $\nabla$, and suitable background
fields $k_R$ and $k_D$. Contrary to the majority of the existing literature on gravitational-wave physics,
we start from a fully nonlinear theory, which is then linearized to explore modified gravitational-wave propagation. Our focus is on the $k_R$ term, whereas a dedicated study of $k_D$ will be left
for a future point in time.

Moreover, we derive modified dispersion relations in the linearized-gravity
regime ourselves without resorting to explicit results in the literature. The first
gravitational-wave event  GW150914~\cite{LIGOScientific:2016aoc} as
well as the binary-neutron star merger GW170817~\cite{LIGOScientific:2017vwq}
in combination with the gamma-ray burst
GRB 170817A~\cite{Goldstein:2017mmi,Savchenko:2017ffs} are employed
to quantify possible deviations from standard linearized gravity that
may be based on the $k_R$ term; see, e.g., Ref.~\cite{LIGOScientific:2017zic}.

Our paper is organized as follows. Section~\ref{sec:modified-gravity-theory}
presents the modified-gravity theory proposed and discusses the properties
essential for us. This theory is linearized in Sec.~\ref{sec:linearization}.
Here, the controlling coefficients are classified according to irreducible
representations of the rotation group. Furthermore, we determine the
modified dispersion relations for gravitational waves for different
sectors of the modification. Section~\ref{sec:phenomenology} is
dedicated to the phenomenological analysis. Both nonbirefringent and
birefringent coefficients are constrained by gravitational-wave
and gamma-ray-burst data. Finally, the results are summarized and discussed in
Sec.~\ref{sec:conclusions}. Natural units are employed with $c=1$
unless stated otherwise. The metric signature is $(-,+,+,+)$. Greek letters are used for spacetime indices, lowercase Latin letters for spatial indices, and uppercase Latin letters for generic indices ranging from 1 to 10.

\section{Spacetime symmetry violation via dim-6 operators}
\label{sec:modified-gravity-theory}

We study a modified-gravity theory perturbed by coordinate-invariant dimension-6 (dim-6) contributions that break diffeomorphism invariance. Explicitly, the extended EH action without cosmological constant takes the following form:
\begin{subequations}\label{eq:action}
\begin{equation}
S=\int_{\mathcal{M}} \mathrm{d}^4x\, \frac{\sqrt{-g}}{2\kappa} \big( R +  \mathcal{L}_g^{(6)} \big)\,,
\end{equation}
where
\begin{align}\label{Lag_R_D}
\mathcal{L}_g^{(6)}&= (k_R^{(6)})^{\alpha\beta\gamma\delta\mu\nu\rho\sigma}
R_{\alpha\beta\gamma\delta}R_{\mu\nu\rho\sigma} \notag \\
&\phantom{{}={}}+(k_D^{(6)})^{\alpha\beta\gamma\delta\kappa\lambda}
\nabla_{(\kappa}\nabla_{\lambda)}R_{\alpha\beta\gamma\delta}\,,
\end{align}
\end{subequations}
with $\kappa=8\pi G_N$ in terms of Newton's gravitational constant $G_N$.
The spacetime manifold $\mathcal{M}$ is
described by the metric tensor $g_{\mu\nu}$ with determinant $g=\det(g_{\mu\nu})$. Moreover, $\nabla_{\mu}$
denotes the $g_{\mu\nu}$-compatible torsion-free covariant derivative. The Riemann curvature tensor on the
spacetime manifold is denoted as $R_{\mu\nu\rho\sigma}$, $R_{\mu\nu}:=R^{\lambda}_{\phantom{\lambda}\mu\lambda\nu}$
indicates the Ricci tensor, and $R:=R^{\lambda}_{\phantom{\lambda}\lambda}$ is the Ricci scalar.
The tensor-valued
background fields $k_R^{(6)}$ and $k_D^{(6)}$ are contracted with dim-6 operators on $\mathcal{M}$, i.e., they
are contained in the nonminimal gravitational SME; cf. Tab.~XVI in Ref.~\cite{Kostelecky:2020hbb}. Note that
we do not consider additional metric, vierbein, or Levi-Civita components, i.e., $k_R^{(6)}$
and $k_D^{(6)}$ are pure background fields. The latter have a generic spacetime dependence
and are taken to be nondynamical, i.e., diffeomorphism invariance is violated explicitly.
When two indices are enclosed by a pair of parentheses, the corresponding expression is symmetrized over these indices. In particular, $X^{(\mu \nu)}:=(X^{\mu \nu}+X^{\nu \mu})/2$ where a normalization factor is included. In principle, proper boundary terms may also be conceived to ensure a well-defined variational principle~\cite{Reyes:2023sgk}. However, they will be omitted for brevity.

The controlling coefficients of $k_R^{(6)}$  and $k_D^{(6)}$ inherit the symmetries of the field operators they are contracted with. For the $k_R^{(6)}$ term, let us group the eight indices in the controlling coefficients into two sets of four indices each. Each set has the symmetries of the Riemann tensor, including the Bianchi-type identities, where the entire eighth-rank tensor is symmetric with respect to interchanging the two sets.
Since the Riemann tensor has 20 independent components, we infer that $k_R^{(6)}$ has $(1/2)\times 20\times 21=210$ independent coefficients.

For the $k_D^{(6)}$ term, the first set of four indices of the background field again has the symmetries of the Riemann tensor. Furthermore, the background is symmetric in the last two indices, which amounts to 10 additional choices for each of the 20 previous possibilities. Interestingly, the second Bianchi identities of pseudo-Riemannian geometry relate certain coefficients to each other. Therefore, the total number of independent coefficients of $k_D^{(6)}$ amounts to 126~\cite{Bailey:2014bta}.

The reasons for examining the dim-6 terms of Eq.~\eqref{eq:action} are multifold. For $d=4$, the only coefficients contributing to gravitational-wave physics are $s^{\mu\nu}$, or, analogously, the double-dual coefficients $s^{(4)\mu\rho\alpha\nu\sigma\beta}$ introduced in Ref.~\cite{Kostelecky:2016kfm}; see also Eq.~(33) in Ref.~\cite{Kostelecky:2016uex}. The $s$ coefficients have already been studied extensively in gravitational-wave physics, see the aforementioned Refs.~\cite{Xu:2019fyt,Xu:2019gua,Xu:2020zxs,Xu:2021dcw,Bailey:2023lzy,Kostelecky:2017zob,Liu:2020slm,Ray:2023sbr,Qin:2023xvz,Hou:2024xbv,AouladLafkih:2025stw,Cao:2024lvd,Rao:2024wde,Liang:2022hxd}, as well as beyond gravitational waves, e.g., in Refs.~\cite{Bailey:2006fd,Iorio:2012gr,Bailey:2013oda,Shao:2014bfa,Kostelecky:2015dpa,Kostelecky:2016uex}. The constant Lorentz scalar $u$ can be absorbed into Newton's constant, and $t^{\mu\nu\rho\sigma}$ only impacts the physics in spacetimes that are not asymptotically flat~\cite{Bonder:2015maa,Bonder:2017dpb}.

Looking into higher-dimensional operators is the next logical step. Additional spacetime derivatives naturally occur in effective field theory with the aim of incorporating physical phenomena at ever smaller length scales. One may be tempted to start with the dim-5 coefficients. However, they are known to exhibit peculiar effects, such as self-acceleration of test bodies in the post-Newtonian regime~\cite{Bailey:2014bta}. Since this behavior raises certain questions about the physical meaning of such modifications, we refrain from examining them in the current paper.

This chain of reasoning then leads us directly to the dim-6 coefficients. It is interesting to note that Eq.~\eqref{eq:action} contains the gravitational
Chern-Simons (CS) term \cite{Jackiw:2003pm} as a special case. Recall that the latter is of the form
\begin{subequations}
\label{eq:CS-theory}
\begin{equation}
S_{\mathrm{CS}}=\frac{1}{2\kappa}\int_{\mathcal{M}} \mathrm{d}^4x\,\frac{1}{4}\theta(x) \,
(\prescript{*}{}{R}R)\,,
\end{equation}
expressed in terms of the gravitational Chern-Pontryagin scalar density
\begin{equation}
\prescript{*}{}{R}R:=\prescript{*}{}{R}^{\sigma\phantom{\tau}
\mu\nu}_{\phantom{\sigma} \tau}R^\tau_{\phantom{\tau} \sigma\mu\nu}\,,
\end{equation}
with the dual of the Riemann curvature tensor
\begin{equation}
{}^{*}R^{\tau\phantom{\sigma}\mu\nu}_{\phantom{\tau}\sigma}:
=\frac{1}{2}\epsilon^{\mu\nu\alpha\beta}R^{\tau}_{\phantom{\tau}\sigma\alpha\beta}\,.
\end{equation}
\end{subequations}
Here, $\theta(x)$ is a dimensionful spacetime-dependent scalar function, and $\epsilon^{\alpha\beta\mu\nu}$ is the four-dimensional totally antisymmetric Levi-Civita symbol. If $\theta$ in Eq.~\eqref{eq:CS-theory} is taken as a nondynamical field, it becomes a source of explicit diffeomorphism violation. Integrating the object $\prescript{*}{}{R}R$ leads to the second Chern number, which renders the CS term topological.

Moreover, Eq.~\eqref{eq:action} contains another topological contribution given by
\begin{subequations}
\label{eq:Einstein-Gauss-Bonnet-theory}
\begin{align}
S_{\mathrm{EGB}}&=\frac{1}{2\kappa}\int_{\mathcal{M}}\mathrm{d}^4x\,({}^{*}R^{*})^{\mu\nu\varrho\sigma}R_{\mu\nu\varrho\sigma}\,, \\[1ex]
({}^{*}R^{*})^{\mu\nu\varrho\sigma}&=\frac{1}{4}\epsilon^{\mu\nu\alpha\beta}\epsilon^{\varrho\sigma\gamma\delta}R_{\alpha\beta\gamma\delta}\,,
\end{align}
\end{subequations}
with the double dual Riemann tensor ${}^{*}R^{*}$. Interestingly, by employing the relationship
\begin{align}
-({}^{*}R^{*})_{\mu\nu\varrho\sigma}&=R_{\mu\nu\varrho\sigma}-(g_{\mu[\varrho}R_{\sigma]\nu}+g_{\nu[\sigma}R_{\varrho]\mu}) \notag \\
&\phantom{{}={}}+\frac{R}{2}g_{\mu[\varrho}g_{\sigma]\nu}\,,
\end{align}
the action of Eq.~\eqref{eq:Einstein-Gauss-Bonnet-theory} is recast into a form expressed in terms of the metric:
\begin{equation}
S_{\mathrm{EGB}}=-16\int\mathrm{d}^4x\,\frac{\sqrt{-g}}{2\kappa}(R^{\mu\nu\varrho\sigma}R_{\mu\nu\varrho\sigma}-4R^{\mu\nu}R_{\mu\nu}+R^2)\,.
\end{equation}
The latter is known as the Einstein-Gauss-Bonnet (EGB) term. It gives rise to a topological quantum number known as the Euler characteristic. We will come back to both of these intriguing terms at a later point in time.

The field equations associated with Eq.~\eqref{eq:action} are obtained by varying the action for the metric.
In the following, we will state the Einstein field equations separately in both sectors. In the $k_R^{(6)}$ sector, for a vanishing $k_D^{(6)}$, they are given by
\begin{align}
\label{eq:modified-einstein-equations}
G^{\mu\nu}&=\frac{1}{2}(k_R^{(6)})^{\alpha\beta\gamma\delta\xi\lambda\rho\sigma} R_{\alpha\beta\gamma\delta}R^\zeta_{\phantom{\nu} \lambda\rho\sigma}\Big(g_{\xi\zeta}g^{\mu\nu}+4\delta_{\phantom{(\mu}\xi}^{(\mu}\delta_{\phantom{\nu)}\zeta}^{\nu)}\Big) \notag \\
&\phantom{{}={}}+4\nabla_{(\sigma}\nabla_{\rho)}\Big((k_R^{(6)})^{\alpha\beta\gamma\delta\nu\sigma\rho\mu}  R_{\alpha\beta\gamma\delta}\Big)\,,
\end{align}
with the Einstein tensor $G^{\mu\nu}:=R^{\mu\nu}-(R/2)g^{\mu\nu}$.
In the $k_D^{(6)}$ sector, we arrive at
\begin{subequations}
\begin{align}
    G^{\mu\nu}&=\frac{1}{2}g^{\mu\nu}(k_D^{(6)})^{\alpha\beta\gamma\delta\kappa\lambda}\nabla_{(\kappa}\nabla_{\lambda)}
    R_{\alpha\beta\gamma\delta}+
      \mathcal T^{\xi\zeta}\delta^{(\mu}_{\phantom{(\mu}\xi} \delta^{\nu)}_{\phantom{\nu)}\zeta}
     \notag \\& \phantom{{}={}} -\frac{1}{2}\nabla_\rho \Big((k_D^{(6)})^{\alpha\beta\gamma\delta\mu\nu}\nabla^\rho
     R_{\alpha\beta\gamma\delta}\Big)\,,
\end{align}
with the second-rank tensor
\begin{align}
\mathcal T^{\xi\zeta}
&=
\nabla_\lambda\nabla_\kappa(k_D^{(6)})^{\xi\beta\gamma\delta\kappa\lambda}R^\zeta{}_{\beta\gamma\delta}
 \notag \\
&\phantom{{}={}}+2\nabla_\beta\nabla_\gamma\nabla_\lambda\nabla_\kappa
(k_D^{(6)})^{\xi\beta\gamma\zeta\kappa\lambda} \notag \\
&\phantom{{}={}}-2\nabla_\alpha \mathcal{J}^{\alpha\beta\gamma\delta\zeta}[R^\xi{}_{\beta\gamma\delta}]
-2\nabla_\lambda\mathcal{J}^{\zeta\beta\gamma\delta\lambda}[R^\xi{}_{\beta\gamma\delta}]
\notag \\
&\phantom{{}={}}+2\nabla_\sigma\mathcal{J}^{\xi\beta\gamma\delta\zeta}[R^{\sigma}_{\phantom{\sigma}\beta\gamma\delta}]
 \notag \\
&\phantom{{}={}}+\nabla_\lambda[(k_D^{(6)})^{\alpha\beta\gamma\delta\zeta\lambda}\nabla^\xi R_{\alpha\beta\gamma\delta}]\,,
\end{align}
expressed in terms of the tensor-valued function
\begin{align}
\mathcal{J}^{A}[X]:=
\nabla_\kappa (k_D^{(6)})^{A\kappa} X
-(k_D^{(6)})^{A\kappa}\nabla_\kappa X \,,
\end{align}
where $\{A\}$ denotes a set of indices.
\end{subequations}
The computer algebra system \textit{Mathematica} in combination with the package \textit{xTras}, which is part of the compilation \textit{xAct} \cite{xTensor:2025}, is a powerful tool for computations of this nature.

\section{Linearization}\label{sec:linearization}
In the following, we linearize the action and the
field equations associated with Eq.~\eqref{eq:action}. To do so, the metric
tensor is decomposed as $g_{\mu\nu}=\eta_{\mu\nu}+h_{\mu\nu}$ with the
Minkowski metric $\eta_{\mu\nu}$ and the perturbation
$\left | h_{\mu\nu} \right| \ll 1$.
From now on, we will refer to the background
fields as simply $k_R$ and $k_D$, respectively, i.e., the label
of the mass dimension will be dropped for brevity.

The linearized action originating from Eq.~\eqref{eq:action} reads
\begin{subequations}
\begin{align}
 S_{\textrm{lin}}&=\frac{1}{2\kappa}\int\mathrm{d}^4x \, \frac{h_{\mu\nu}}{4}\Big(\hat{G}^{\mu\nu\rho\sigma}+\hat{M}_R^{\mu\nu\rho\sigma}+\hat{M}_D^{\mu\nu\rho\sigma}    \Big)h_{\rho\sigma}   \,,
\end{align}
where we introduced the following short-hand notation for the standard contributions,
\begin{align}
\hat{G}^{\mu\nu\rho\sigma}&= -\big(\eta^{\nu\rho}\partial^\mu\partial^\sigma +\eta^{\mu\sigma}\partial^\nu \partial^\rho \big)+\eta^{\rho\sigma}\partial^\mu\partial^\nu\notag \\
&\phantom{{}={}}+\eta^{\mu\nu}\partial^\rho\partial^\sigma -\eta^{\mu\nu}\eta^{\rho\sigma}\Box + \eta^{\mu\rho}\eta^{\nu\sigma}\Box  \,,
\end{align}
and the effective modifications
\begin{align}
\hat{M}_R^{\mu\nu\rho\sigma}&=
16(k_R)^{\mu\alpha\nu\beta\rho\gamma\sigma\delta}\partial_\alpha\partial_\beta
\partial_\gamma\partial_\delta   \,,
\\[1ex]
\hat{M}_D^{\mu\nu\rho\sigma}&=4\big[(k_D)^{\rho\alpha\sigma\beta\mu\nu}
\Box  \partial_\alpha \partial_\beta \notag \\
&\phantom{{}={}}\quad- 2 (k_D)^{(\mu\alpha\rho\beta\kappa\lambda} \eta^{\nu)\sigma}\partial_\kappa\partial_\lambda   \partial_\alpha \partial_\beta \notag \\
&\phantom{{}={}}\quad +2 (k_D)^{(\mu\rho\sigma\alpha\kappa\lambda} \partial^{\nu)}\partial_\kappa\partial_\lambda  \partial_\alpha \notag \\
&\phantom{{}={}}\quad +2(k_D)^{(\mu\alpha\rho\beta\nu)\lambda} \partial_\lambda   \partial_\alpha  \partial_\beta \partial^\sigma  \notag \\
&\phantom{{}={}}\quad -2(k_D)^{(\mu\rho\sigma\alpha\nu)\lambda} \Box\partial_\lambda  \partial_\alpha   \notag \\
&\phantom{{}={}}\quad -4(k_D)^{\rho\alpha\sigma\beta(\mu\lambda} \partial^{\nu)}\partial_\lambda \partial_\alpha \partial_\beta  \notag \\
&\phantom{{}={}}\quad -\eta^{\mu\nu}(k_D)^{\rho\alpha\sigma\beta\kappa\lambda}  \partial_\kappa\partial_\lambda  \partial_\alpha \partial_\beta \big]\,.
\end{align}
\end{subequations}
The complete linearized field equations are then
\begin{subequations}
\label{eq:linearized-field-equations}
\begin{equation}
\hat{L}^{\mu\nu}+\hat{Q}_R^{\mu\nu}+\hat{Q}_D^{\mu\nu}=0\,,
\end{equation}
with the Lichnerowicz operator
\begin{align}
\label{eq:lichnerowicz-operator}
\hat{L}^{\mu\nu}&=-\square h^{\mu\nu}+2\partial_{\rho}\partial^{(\mu}h^{\nu)\rho}-\partial^{\mu}\partial^{\nu}h \notag \\
&\phantom{{}={}}+(\square h-\partial^{\rho}\partial^{\sigma}h_{\rho\sigma})\eta^{\mu\nu}\,,
\end{align}
the contribution associated with $k_R$,
\begin{equation}
\hat{Q}_R^{\mu\nu}=-16 (k_R)^{\mu\alpha\nu\beta\kappa\gamma\lambda\delta}   \partial_\alpha \partial_\beta \partial_\gamma \partial_\delta h_{\kappa\lambda} \,,
\end{equation}
as well as another term depending on $k_D$:
\begin{align}
\hat{Q}_D^{\mu\nu}&=4\big[-(k_D)^{\alpha\beta\gamma\delta\mu\nu}\Box  \partial_\delta \partial_\beta h_{\gamma\alpha} \notag \\
&\phantom{{}={}}\quad + 2 (k_D)^{(\mu\alpha\beta\gamma\kappa\lambda}\partial_\kappa\partial_\lambda  \partial_\gamma\partial_\alpha h_{\beta}^{\phantom{\gamma}\nu)}\notag \\
&\phantom{{}={}}\quad -2 (k_D)^{(\mu\alpha\beta\gamma\kappa\lambda}\partial^{\nu)}\partial_\kappa\partial_\lambda\partial_\gamma  h_{\alpha\beta} \notag \\
&\phantom{{}={}}\quad - 2(k_D)^{(\mu\beta\gamma\delta\nu)\lambda}\partial_\lambda  \partial_\delta \partial_\beta\partial_\alpha h_{\gamma}^{\phantom{\gamma}\alpha} \notag \\
&\phantom{{}={}}\quad +2(k_D)^{(\mu\alpha\beta\gamma\nu)\lambda}\Box\partial_\lambda  \partial_\gamma  h_{\alpha\beta}\notag \\
&\phantom{{}={}}\quad +4(k_D)^{\alpha\beta\gamma\delta(\mu\lambda}\partial^{\nu)}\partial_\lambda \partial_\delta\partial_\beta h_{\gamma\alpha}\notag \\
&\phantom{{}={}}\quad +\eta^{\mu\nu}(k_D)^{\alpha\beta\gamma\delta\kappa\lambda}  \partial_\kappa\partial_\lambda\partial_\beta\partial_\delta h_{\alpha\gamma}\big]\,,
\end{align}
\end{subequations}
where $h:=h^{\mu}_{\phantom{\mu}\mu}$ is the trace of the metric perturbation.
It is interesting to note that the operators are transverse and therefore gauge-invariant~\cite{Kostelecky:2016kfm,Kostelecky:2017zob}.

In Minkowski spacetime, SME coefficients are usually taken as spacetime constants to preserve translation invariance and energy-momentum conservation. Nevertheless, extensions involving spacetime-dependent coefficients cannot be ruled out \textit{a priori}~\cite{Lane:2016osk,Riquelme:2026ztl,CesareSilva:2026zba}.

In curved spacetimes, the tensor fields $k_R$ and $k_D$ have a spacetime dependence that follows from solving the modified Einstein equations, which is a highly challenging endeavor. In a gravitational context, it is impossible to choose SME coefficients as constants. Requirements of the form $\partial_{\mu}k_{R,D}=0$ are meaningless since the partial derivative does not transform as a four-vector, which makes such a condition dependent on the choice of coordinates. One may be tempted to propose the alternative conditions $\nabla_{\mu}k_{R,D}=0$, which, for a generic spacetime, do not possess nontrivial solutions, though \cite{Kostelecky:2003fs}.

When linearizing gravity to study gravitational-wave phenomena, general coordinate transformations and diffeomorphisms are now linearized too, and they can be interpreted as gauge transformations~\cite{Kostelecky:2020hbb}. It is then a mild and reasonable assumption to consider spacetime-constant tensor-valued SME background fields. From a physical viewpoint, the spacetime dependence of $k_{R,D}$ is expected to be governed by gravitational fields of large and intermediate scales, such as those of the Milky Way and the Earth, respectively, which are comparatively weak. After all, linearization is performed around Minkowski spacetime. Moreover, the background field is not supposed to be significantly modulated by the extremely tiny ripple effects of spacetime that a gravitational wave poses.

Obviously, although linearized, the field equations of $k_D$ retain a certain complexity in contrast to those of $k_R$, which exhibit a single additional term besides the Lichnerowicz operator of Eq.~\eqref{eq:lichnerowicz-operator}.
For this reason, a dedicated study of the $k_D$ term is challenging from a technical standpoint. Thus, it is left for a future point in time, and for the remainder of the paper, our focus will be on the $k_R$ term.

Interestingly, Eq.~\eqref{eq:linearized-field-equations} for $k_R$ shares certain characteristics with the field equations of the CPT-even part of the electromagnetic sector in the SME~\cite{Colladay:1998fq}, when the latter is expressed via the electromagnetic vector potential. In principle, Eq.~\eqref{eq:linearized-field-equations} contains two copies of the modified inhomogeneous Maxwell equations without a source. Note that the relationship between standard linearized gravity and \textit{U}(1) gauge theory has been pointed out before; see Ref.~\cite{Bern:2002kj}. This finding is expected to apply to the gravitational SME, as long as the investigation is restricted to gauge-invariant operators. In a forthcoming chain of reasoning, we will benefit from close parallels between electromagnetism and linearized gravity.

It makes sense to express Eq.~\eqref{eq:linearized-field-equations} in terms of the trace-reversed metric perturbation
\begin{equation}
\label{eq:trace-reversed-variables}
h_{\mu\nu}:=\bar{h}_{\mu\nu}-\frac{\bar{h}}{2}\eta_{\mu\nu}\,,\quad h=-\bar{h}\,,
\end{equation}
which simplifies the standard part of Eq.~\eqref{eq:linearized-field-equations} significantly. By employing the Lorenz gauge $\partial_{\mu}\overline{h}^{\mu\nu}=0$ and setting $k_D=0$, we arrive at
\begin{align}
\label{eq:linearized-field-equations-lorentz-gauge}
0&=-\Box \bar{h}^{\mu\nu}-16(k_R)^{\alpha\beta\gamma\delta\nu\sigma\rho\mu}\partial_\delta \partial_\alpha\partial_\sigma \partial_\rho \notag \\ &\phantom{{}={}} \times \bigg(\delta_{\phantom{\kappa}\beta}^\kappa \delta_{\phantom{\lambda}\gamma}^\lambda-\frac{1}{2}\eta_{\beta\gamma}\eta^{\kappa\lambda}\bigg)\bar{h}_{\kappa\lambda}\,.
\end{align}
These field equations describe gravitational-wave propagation in a
linearized gravity theory based on Eq.~\eqref{eq:action} with $k_D$ discarded.
Equation~\eqref{eq:linearized-field-equations-lorentz-gauge} is a set of modified
wave equations for the independent components of the trace-reversed metric
perturbation that makes up the gravitational wave when propagating in spacetime.
The latter equations must be solved to understand how a gravitational wave is
affected by the background field~$k_R$.

We transform the field equations to the frequency domain and express them in the following form:
\begin{subequations}
\label{eq:linearized-field-equations-momentum-space}
\begin{equation}
0=\bar{M}^{\mu\nu\kappa\lambda}(p)\overline{h}_{\kappa\lambda}(p)\,,
\end{equation}
with the fourth-rank tensor
\begin{align}
\bar{M}^{\mu\nu\kappa\lambda}(p)&=p^2\Omega^{\mu\nu\kappa\lambda}-16\mathcal{S}^{\mu\nu\kappa\lambda}\,,
\end{align}
and the definitions
\begin{align}
\Omega^{\mu\nu\kappa\lambda}&:=\frac{1}{2}(\eta^{\kappa\mu}\eta^{\lambda\nu}
+\eta^{\lambda\mu}\eta^{\kappa\nu})\,, \displaybreak[0]\\[2ex]
\mathcal{S}^{\mu\nu\kappa\lambda}&:=
\tilde{K}^{\beta\circ\gamma\circ\mu
\circ\nu\circ}\left(\delta^{\kappa}_{\phantom{\kappa}\beta}
\delta^{\lambda}_{\phantom{\lambda}\gamma}-\frac{1}{2}
\eta_{\beta\gamma}\eta^{\kappa\lambda}\right)\,,
\end{align}
\end{subequations}
where $\overline{h}_{\mu\nu}(p)$ is the Fourier-transformed
trace-reversed metric perturbation dependent on the wave four-vector~$p_{\mu}$. Here we introduced the
short-hand notation of coefficients contracted with various wave four-vectors:
\begin{equation}
\label{eq:definition-K-tilde}
\tilde{K}^{\beta\circ\gamma\circ\mu\circ\nu\circ}:=
(k_R)^{\beta\alpha\gamma\delta\mu\rho\nu\sigma}p_{\sigma}p_{\rho}p_{\delta}p_{\alpha}\,,
\end{equation}
where each contraction is indicated by the symbol `$\circ$.'
This notation is adopted from Ref.~\cite{Kostelecky:2017zob}.

The latter form of the wave equations is valuable since they bear similarity to the wave equations for the electromagnetic vector
potential $A_{\mu}$, which read $0=M^{\mu\nu}A_{\nu}$. Here, $M^{\mu\nu}$ is a suitable
$(4\times 4)$ matrix that follows from the Maxwell equations.
Now, Eq.~\eqref{eq:linearized-field-equations-momentum-space} is the analog
of the electromagnetic field equations for the trace-reversed metric perturbation, which is a spin-2
field. Note that $\bar{M}^{\mu\nu\kappa\lambda}$
has been symmetrized in the first and
second pairs of indices, respectively.
However, $\bar{M}^{\mu\nu\kappa\lambda}$ should
not be assumed symmetric with respect to interchanging both index pairs.

There is a trick for how to reduce $\bar{M}^{\mu\nu\kappa\lambda}$
to a second-rank tensor, which is represented by a $(10\times 10)$ matrix.
We introduce indices $A,B\in \{(0,0):=1,(0,1):=2,(0,2):=3,\dots,(3,3):=10\}$ whose values
correspond to one of the 10 possibilities of ordered pairs
of the numbers $0,1,2,3$. Then,
\begin{subequations}
\label{eq:free-field-equations-reduced}
\begin{align}
0&=\bar{M}^{AB}(p)\overline{h}_B(p)\,, \\[2ex]
\bar{M}^{AB}(p)&=p^2\Omega^{AB}-16\mathcal{S}^{AB}(p)\,.
\end{align}
\end{subequations}
These free-field equations have nontrivial solutions for the metric
perturbation when the determinant of the coefficient matrix vanishes; see also Ref.~\cite{Bailey:2024zgr}.
An expansion of the determinant to second order in the controlling
coefficients is indispensable to obtain the correct dispersion relations
at first order in the controlling coefficients.

\subsection{Number of independent component coefficients}
\label{sec:number-coefficients}

The current paper is one of the few in contemporary literature that starts from a modification of the EH action instead of an alteration of the action at the linearized level. Therefore, differences compared to the latter approach are expected. One of these concerns the number of independent component coefficients of $k_R$, which amounts to 210 due to arguments about index symmetries. Now, according to Tab.~I in Ref.~\cite{Kostelecky:2017zob}, there is actually no configuration of coefficients for $d=6$ that exhibits this very same number. Furthermore, none of the Young diagrams has the same symmetry properties as $k_R$, which raises questions.

This apparent issue is understood when noting that the authors of Ref.~\cite{Kostelecky:2017zob} construct all coefficient sets transforming under irreducible representations of $\mathit{SO}(1,3)$, i.e., their Tab.~I is exhaustive in this respect. However, the direct product of two Riemann tensors transforms under the \textit{reducible} representation given by
\begin{equation}
\yng(2,2)\otimes \yng(2,2)\,,
\end{equation}
where $\otimes$ stands for the direct product, and each set of four boxes describes the irreducible representation under which the Riemann tensor transforms. This product representation must be decomposed into a direct sum of irreducible representations of $\mathit{SO}(1,3)$. By explicitly requiring that the product be symmetric with respect to interchanging both Young diagrams, we arrive at the decomposition:
\begin{align}
\label{eq:decomposition-product-representation}
\yng(2,2)\otimes_s \yng(2,2)&=\yng(4,4)\, \oplus\, \yng(4,2,2)\nonumber \\
&\phantom{{}={}}\oplus\, \yng(3,3,1,1)\, \oplus\, \yng(2,2,2,2)\,,
\end{align}
with the symmetrized direct product $\otimes_s$ and the direct sum $\oplus$ of representations. The first Young diagram on the right-hand side stands for the product of two Weyl tensors, and the last describes the single configuration that gives rise to the already discussed EGB term of Eq.~\eqref{eq:Einstein-Gauss-Bonnet-theory}. By evaluating the dimensions of these representations with the help of the hook content and hook length formulas~\cite{Hamermesh:2026}, we obtain
\begin{equation}
\frac{1}{2}\times 20\times 21=210=105+84+20+1\,,
\end{equation}
which is consistent. In fact, the first summand on the right-hand side corresponds to the number of component coefficients for $k^{(6)}$ in the tenth line Tab.~I of Ref.~\cite{Kostelecky:2017zob}. The second summand is the number of component coefficients of $s^{(6)}$ in the first line of the very same table. These findings already show how coefficient sets transforming under distinct irreducible representations of $\mathit{SO}(1,3)$ can mix with each other when starting from a nonlinear theory.

The form of the fourth Young diagram on the right-hand side of Eq.~\eqref{eq:decomposition-product-representation} suggests that the set of component coefficients transforming under this representation must involve two Levi-Civita tensors, i.e., they are taken as $(k_R)^{\alpha\beta\gamma\delta\nu\rho\sigma\mu}=\epsilon^{\alpha\beta\nu\rho}\epsilon^{\gamma\delta\sigma\mu}/(4\sqrt{-g})$. Indeed, this choice implies the EGB term given by Eq.~\eqref{eq:Einstein-Gauss-Bonnet-theory}.

The third and fourth Young diagrams do not contribute at the linearized level. After all, coefficient sets with these symmetries are eliminated when contracted with four partial derivatives or wave four-vectors. This holds, in particular, for the EGB term, which depends on a single coefficient and is associated with the last Young diagram on the right-hand side of Eq.~\eqref{eq:decomposition-product-representation}. The EGB action is a topological invariant and does not contribute to the field equations.

The situation is similar for the gravitational CS term, which depends on a single Levi-Civita symbol. As was figured out before, the latter does not modify gravitational-wave propagation at the kinematic level~\cite{Jackiw:2003pm}, although it can still play some role beyond kinematics. An immediate implication is that some of the original 210 coefficients are no longer independent of one another, which leads to certain degeneracies. Note that Ref.~\cite{Chung:2022ees} shows how to perform a systematic decomposition of direct products of the Riemann tensor into objects that transform under irreducible representations, which also includes how to count the number of independent components. While we do not make use of these results, they are expected to be valuable for more complicated terms.

The modified field equations~\eqref{eq:modified-einstein-equations} reveal another significant point to take into account. The chain of reasoning is easier to follow in Maxwell electrodynamics and modifications thereof in the context of the minimal SME. Thus, we will first dedicate ourselves to a thorough description of the situation in electrodynamics, before delving into modified linearized gravity. Although the arguments in both settings are analogous, linearized gravity is accompanied by additional technical challenges.

\subsubsection{Bianchi identities in electromagnetism}

Maxwell electrodynamics is a $\mathit{U}(1)$ gauge theory with connection $A_{\mu}$. The curvature on the principal bundle is the electromagnetic field strength $F_{\mu\nu}$. The homogeneous Maxwell equations arise as Bianchi identities on the principal bundle. They can be expressed either via a cyclic sum,
\begin{subequations}
\label{eq:maxwell-equations-homogeneous}
\begin{equation}
\sum_{(\mu\nu\varrho)} \partial_{\mu}F_{\nu\varrho}=0\,,
\end{equation}
or in terms of the dual field strength $\tilde{F}^{\mu\nu}:=\frac{1}{2}\epsilon^{\mu\nu\varrho\sigma}F_{\varrho\sigma}$ such that
\begin{equation}
\partial_{\nu}\tilde{F}^{\nu\mu}=0=\epsilon^{\mu\nu\varrho\sigma}\partial_{\nu}F_{\varrho\sigma}\,.
\end{equation}
\end{subequations}
Since the SME maintains gauge invariance, it preserves the properties of the principal bundle, i.e., the homogeneous Maxwell equations are taken over unchanged. What changes, though, are the inhomogeneous Maxwell equations \textit{in vacuo}, which in the minimal SME take the form
\begin{equation}
\label{eq:inhomogeneous-maxwell-SME}
\partial_{\nu}F^{\mu\nu}+k^{\mu\nu\varrho\sigma}\partial_{\nu}F_{\varrho\sigma}=0\,,
\end{equation}
with a generic background field $k$, which transforms as a fourth-rank Lorentz tensor under boosts and rotations of the coordinate frame. Interestingly, applying the homogeneous Maxwell equations to Eq.~\eqref{eq:inhomogeneous-maxwell-SME} eliminates a totally antisymmetric $k$:
\begin{subequations}
\label{eq:introduction-kF}
\begin{align}
0&=\partial_{\nu}F^{\mu\nu}+(k_F)^{\mu\nu\varrho\sigma}\partial_{\nu}F_{\varrho\sigma}\,, \\[1ex]
(k_F)^{\mu\nu\varrho\sigma}&:=k^{\mu\nu\varrho\sigma}-\theta\epsilon^{\mu\nu\varrho\sigma}\,,
\end{align}
\end{subequations}
where $\theta$ is a real, constant Lorentz scalar. The components of the new background field $k_F$ satisfy a Bianchi-type identity:
\begin{equation}
\label{eq:bianchi-identity-electromagnetic-coefficients}
\sum_{(\nu\varrho\sigma)} (k_F)^{\mu\nu\varrho\sigma}=0\,.
\end{equation}
Hence, requiring Eq.~\eqref{eq:bianchi-identity-electromagnetic-coefficients} is equivalent to eliminating the completely antisymmetric piece of the background field~$k$. The piece referred to is known as the $\mathit{U}(1)$ $\theta$ term given by $\theta F_{\mu\nu}\tilde{F}^{\mu\nu}=-4\theta\mathbf{E}\cdot\mathbf{B}$, which arises from the choice $k^{\mu\nu\varrho\sigma}=\theta\epsilon^{\mu\nu\varrho\sigma}$. Here, $\mathbf{E}$ and $\mathbf{B}$ are the electric and magnetic fields, respectively. The $\theta$ term is a mere surface term and does not contribute to the field equations.

However, replacing $\theta$ by a coordinate-dependent field $\theta(x)$ implies the well-known axion interaction~\cite{Wilczek:1987mv}, which is more than just a surface term. Furthermore, integrating the latter by parts provides the Carroll-Field-Jackiw (CFJ) term $(k_{AF})_{\mu}A_{\nu}\tilde{F}^{\mu\nu}$ with a vector-valued background field $k_{AF}$, where $(k_{AF})_{\mu}=\partial_{\mu}\theta$~\cite{Carroll:1989vb}. Note that the homogeneous Maxwell equations do not get rid of the CFJ term, which shows that it is physical.

The entire argument is contained in the irreducible $\mathit{SO}(1,3)$ representations that the field strength and background fields transform under. The generic background field $k$ introduced in Eq.~\eqref{eq:inhomogeneous-maxwell-SME} transforms under a \textit{reducible} product of representations for the field strength, which can be rewritten as a direct sum over irreducible $\mathit{SO}(1,3)$ representations:
\begin{equation}
\label{eq:em-direct-sum-irreps}
\yng(1,1)\otimes_s \yng(1,1)=\yng(2,2)\oplus \yng(1,1,1,1)\,.
\end{equation}
Consulting the dimensions of the representations, we have $\frac{1}{2}\times 6\times 7=20+1$. The object $k_F$ introduced in Eq.~\eqref{eq:introduction-kF} transforms under the first representation on the right-hand side and has 20 components, whereas the single component $\theta\epsilon^{\mu\nu\varrho\sigma}$ transforms under the second representation.

The second Bianchi identities mandate that there be no configuration of $k$ that transforms under the representation of the second Young diagram on the right-hand side of Eq.~\eqref{eq:em-direct-sum-irreps}. That is desirable for constant $\theta$, as it removes all unphysical terms from the theory. However, when $\theta=\theta(x)$, the homogeneous Maxwell equations cannot be employed to eliminate this term, since it is physical.

SME coefficients are considered equivalent when they lead to the same field equations, e.g., coefficients related by index symmetries. As long as $\theta$ is constant, Eq.~\eqref{eq:bianchi-identity-electromagnetic-coefficients} increases the number of members of the equivalence classes of coefficients. The observable coefficient space is given by the total coefficient space modulo the dependent coefficients via index symmetries and Eq.~\eqref{eq:bianchi-identity-electromagnetic-coefficients}. In order words, the theory has superfluous degrees of freedom that should be eliminated based on Eq.~\eqref{eq:bianchi-identity-electromagnetic-coefficients}.

An illustrative demonstration of this procedure is as follows. Equation~\eqref{eq:bianchi-identity-electromagnetic-coefficients} is automatically satisfied for almost all index sets due to the index symmetries of $k_F$ according to the first Young diagram on the right-hand side of Eq.~\eqref{eq:em-direct-sum-irreps}. An exception is the choice $\{0,1,2,3\}$. In that case, the SME modification of Eq.~\eqref{eq:inhomogeneous-maxwell-SME} explicitly reads
\begin{align}
-k^{\mu\nu\varrho\sigma}\partial_{\nu}F_{\varrho\sigma}&=k^{0123}\partial_1B^1+k^{0231}\partial_2B^2 \notag \\
&\phantom{{}={}}+k^{0312}\partial_3B^3\,,
\end{align}
with the components $B^i$ of the magnetic field. By subtractig a totally antisymmetric piece $\Lambda\epsilon^{\mu\nu\varrho\sigma}$ with a real constant~$\Lambda$,
\begin{align}
-(k^{\mu\nu\varrho\sigma}+\Lambda\epsilon^{\mu\nu\varrho\sigma})\partial_{\nu}F_{\varrho\sigma}&=(k^{0123}+\Lambda)\partial_1B^1 \notag \\
&\phantom{{}={}}+(k^{0231}+\Lambda)\partial_2B^2 \notag \\
&\phantom{{}={}}+(k^{0312}+\Lambda)\partial_3B^3\,.
\end{align}
The Gauss law $\boldsymbol{\nabla}\cdot\mathbf{B}=0$ for the magnetic field conveys that $\Lambda$ is not observable. Consequently, the observable coefficient space is actually formed by only 2 of the 3 coefficients with indices $\{0,1,2,3\}$. A possible choice is $\Lambda=-k^{0312}$, which leads to the special observable representatives $k^{0123}-k^{0312}$ and $k^{0231}-k^{0312}$. Alternatively, we can work with $k_F$ subject to Eq.~\eqref{eq:bianchi-identity-electromagnetic-coefficients}, which is fulfilled, e.g., by imposing the relationship $(k_F)^{0312}=-(k_F)^{0123}-(k_F)^{0231}$ between the 3 coefficients. The observable representatives are then $(k_F)^{0123}$ and $(k_F)^{0231}$.

\subsubsection{Differential Bianchi identities in linearized gravity}
\label{eq:differential-bianchi-identities}

The previous arguments are now adopted to the setting of modified linearized gravity studied. The spacetime manifold $\mathcal{M}$ has curvature $R_{\mu\nu\varrho\sigma}$. The differential Bianchi identities for the Riemann tensor, referred to as the second Bianchi identities in the following, are the analog of the homogeneous Maxwell equations in electrodynamics. At first order in the metric perturbation, the last term on the right-hand side of Eq.~\eqref{eq:modified-einstein-equations} is the dominant one. The linearized second Bianchi identities then allow us to reformulate the latter:
\begin{align}
G^{(1)\mu\nu}&=4(k_R)^{\alpha\beta\gamma\delta\nu\sigma\rho\mu}\partial_{(\sigma}\partial_{\rho)}R^{(1)}_{\alpha\beta\gamma\delta} \notag \\
&=4(k_R)^{\alpha\beta\gamma\delta\nu\sigma\rho\mu}\partial_{\sigma}\partial_{\rho}R^{(1)}_{\alpha\beta\gamma\delta} \notag \\
&=-4(k_R)^{\alpha\beta\gamma\delta\nu\sigma\rho\mu}\partial_{\sigma}(\partial_{\delta}R^{(1)}_{\alpha\beta\rho\gamma}+\partial_{\gamma}R^{(1)}_{\alpha\beta\delta\rho}) \notag \\
&=-4\Big[(k_R)^{\alpha\beta\delta\rho\nu\sigma\gamma\mu}+(k_R)^{\alpha\beta\rho\gamma\nu\sigma\delta\mu}\Big]\partial_{\sigma}\partial_{\rho}R^{(1)}_{\alpha\beta\gamma\delta}\,,
\end{align}
with the linearized Einstein and Riemann curvature tensors $G^{(1)\mu\nu}$ and $R^{(1)}_{\alpha\beta\gamma\delta}$, respectively, whose explicit forms are not essential for the argument. Then, the coefficients are subject to the restriction
\begin{equation}
\Big[\sum_{(\gamma\delta\varrho)} (k_R)^{\alpha\beta\gamma\delta\nu\sigma\rho\mu}\Big]\partial_{\sigma}=0\,,
\end{equation}
in position space and
\begin{subequations}
\label{eq:contracted-bianchi-identity-coefficients}
\begin{align}
\mathcal{K}^{\alpha\beta\gamma\delta\nu\sigma\rho\mu}p_{\sigma}&=0\,, \\[1ex]
\mathcal{K}^{\alpha\beta\gamma\delta\nu\sigma\rho\mu}&:=\sum_{(\gamma\delta\varrho)} (k_R)^{\alpha\beta\gamma\delta\nu\sigma\rho\mu}\,,
\end{align}
\end{subequations}
in the frequency domain, respectively. Note the additional derivative (or wave vector), which is not present in electromagnetism due to dimensional reasons and since linearized gravity involves two $\textit{U}(1)$ structures.

Equation~\eqref{eq:contracted-bianchi-identity-coefficients} is a contracted Bianchi-type identity for~$k_R$. In principle, $\mathcal{K}$ describes a multilinear map, which is applied to the wave four-vector. When the $p_0$ component occurs, the kernel of this map is empty, since the gravitational wave is on-shell, and Eq.~\eqref{eq:contracted-bianchi-identity-coefficients} does not correspond to the physical dispersion equation. If $p_0$ does not occur, the kernel may contain discrete spatial wave vectors $\mathbf{p}^{(1)},\mathbf{p}^{(2)},\dots$, but each one of these poses a set of measure zero. Thus, for Eq.~\eqref{eq:contracted-bianchi-identity-coefficients} to be satisfied for arbitrary $p_{\sigma}$, the $k_R$ components must obey the following Bianchi-type identity:
\begin{equation}
\label{eq:bianchi-identity-coefficients}
\sum_{(\gamma\delta\varrho)} (k_R)^{\alpha\beta\gamma\delta\nu\sigma\rho\mu}=0\,.
\end{equation}
In a manner analogous to Eq.~\eqref{eq:bianchi-identity-electromagnetic-coefficients}, the second Bianchi identities of the curvature act as a filter for coefficients that do not obey Eq.~\eqref{eq:bianchi-identity-coefficients}. It increases the number of members in equivalence classes of coefficients, i.e., it reduces the number of observable coefficients.

To understand the implications of Eq.~\eqref{eq:bianchi-identity-coefficients} better, we define index pairs of E, B, and G-type in analogy to how the electromagnetic fields $\mathbf{E}$ and $\mathbf{B}$ are extracted from the antisymmetric field strength tensor. Each of these labels refers to one of the two four-index sets in the controlling coefficients $k_R$. It is understood that $e^i:=(0,i)\in \{(0,1);(0,2);(0,3)\}$ and $b^i:=\epsilon^{ijk}(j,k)\in \{(2,3);(1,3);(1,2)\}$ where $\epsilon^{ijk}$ is the totally antisymmetric Levi-Civita symbol in three dimensions.

When each four-index block of $k_R$ has E-type indices, we deduce that
\begin{subequations}
\label{eq:EB-next-EB}
\begin{align}
\label{eq:E-next-E}
(k_R)^{\dots e^ie^j\dots}&=(k_R)^{\dots 0i0j\dots}=-(k_R)^{\dots 00ji\dots}-(k_R)^{\dots 0ji0\dots} \notag \\
&=(k_R)^{\dots 0j0i\dots}\,,
\end{align}
and similarly for B-type indices,
\begin{align}
\label{eq:B-next-B}
(k_R)^{\dots b^kb^j\dots}&=(k_R)^{\dots ijik\dots}=-(k_R)^{\dots iikj\dots}-(k_R)^{\dots ikji\dots} \notag \\
&=(k_R)^{\dots ikij\dots}\,.
\end{align}
\end{subequations}
If an E-type neighbors a B-type index, we have
\begin{subequations}
\label{eq:EB-identity}
\begin{equation}
(k_R)^{\dots 0ijk\dots}+(k_R)^{\dots 0jki\dots}+(k_R)^{\dots 0kij\dots}=0\,,
\end{equation}
which can be interpreted as
\begin{equation}
\sum_{i=1\dots 3} (k_R)^{\dots e^ib^i\dots}=0\,.
\end{equation}
\end{subequations}
By exploiting the other symmetries of $k_R$, the latter relationship is reformulated, and we arrive at the following finding:
\begin{align}
\label{eq:E-next-B}
(k_R)^{\dots 0ijk\dots}&=-(k_R)^{\dots i0jk\dots}=(k_R)^{\dots ijk0\dots}+(k_R)^{\dots ik0j\dots} \notag \\
&=-(k_R)^{\dots ij0k\dots}-(k_R)^{\dots ki0j\dots} \notag \\
&=-(k_R)^{\dots jk0i\dots}=(k_R)^{\dots jk0i\dots}\,.
\end{align}
Therefore, if the second Bianchi identities are used, the eighth-rank tensor $k_R$ becomes completely symmetric in the neighboring pairs of indices. Consequently, its $s^{(6)}$ part is eliminated, and what remains are the coefficients contained in $k^{(6)}$, of which there are 105.
Recall the analogous observation in the electromagnetic sector of the minimal SME, where the $\theta$ term with constant $\theta$ is eliminated by the homogeneous Maxwell equations; cf. Eq.~\eqref{eq:em-direct-sum-irreps}.

The condition of Eq.~\eqref{eq:bianchi-identity-coefficients} gets rid of $k_R$ configurations that are completely antisymmetric over sets of more than 2 indices. This refers to all but the first Young diagram on the right-hand side of Eq.~\eqref{eq:decomposition-product-representation}. Examples are the
gravitational CS term and the EGB term. Indeed, these are surface terms for constant coefficients, such that they do not contribute to gravitational-wave dynamics. In analogy to the $\theta$ term in electrodynamics, which couples the electric to the magnetic field in the Lagrange density, the gravitational CS and EGB terms couple E- and B-type indices with each other. The differential Bianchi identities of electromagnetism and linearized gravity, respectively, obliterate such couplings.

However, as in the electromagnetic case, there are obstructions to using the second Bianchi identities. Imagine that the gravitational CS and EGB terms are coupled to scalar spacetime-dependent fields. Then, suitable integrations by parts imply CFJ-like contributions that are physical; see, e.g., Ref.~\cite{Jackiw:2003pm}.

To summarize, Eq.~\eqref{eq:bianchi-identity-coefficients} eliminates all uninteresting surface terms that may be present for spacetime-constant coefficients. However, these terms become significant when coupled to spacetime-dependent scalar fields. The latter provide additional degrees of freedom necessary to give physical significance to contributions that are uninteresting otherwise. Thus, the condition restricts $k_R$ to its $k_F$-like part. Doing so omits potential $k_{AF}$-like terms, which only arise for coordinate-dependent coefficients.

Contrary to electrodynamics, there are many more of these terms due to the significantly larger number of components of $k_R$, as compared to $k$ of Eq.~\eqref{eq:inhomogeneous-maxwell-SME}. We will refer to such cases throughout the remainder of the paper. A dedicated study of these terms poses an interesting project, but is not pursued here.

Last but not least, note also that the current consideration is valid at leading order in the metric perturbation. Hence, even if a certain configuration of coefficients does not lead to physical effects at the linearized level, this does not say anything about its potential impact in the nonperturbative regime of gravity, such as for black-hole studies.

\subsection{Irreducible coefficient sets}

Due to the generally large number of independent component coefficients, it
is valuable to decompose $k_R$ into subsets that transform under irreducible
representations of the three-dimensional
rotation group $\mathit{SO}(3)$; see Tab.~\ref{tab:subsets-coefficients}. First of all, this is accomplished for the original 210 component coefficients of $k_R$ without considering Eq.~\eqref{eq:bianchi-identity-coefficients}.

We proceed as follows and make use of the terminology introduced previously. Considering the first four indices, coefficients of the form $(k_R)^{0i0j\dots}$, $\epsilon^{jmn}(k_R)^{0imn\dots}$, and $\epsilon^{ikl}\epsilon^{jmn}(k_R)^{klmn\dots}$ are denoted as E-type, G-type and B-type, respectively. Analogous terminology is employed for the second set of four indices. The six possibilities, together with the number of component coefficients deduced from index symmetries, are compiled in Tab.~\ref{tab:subsets-coefficients}.

These sets transform under product representations of $\mathit{SO}(3)$. The second step is to employ group theory arguments to decompose the latter into direct sums of irreducible representations of
$\mathit{SO}(3)$. Let us follow the ordering of Tab.~\ref{tab:subsets-coefficients}. To work out the decompositions, the symmetry and tracelessness properties of the six sets of coefficients are indispensable, and these are provided in Tab.~\ref{tab:properties-coefficients}.

Each block of four indices transforms under a \textit{reducible} representation since the fourth-rank tensor can be understood as a product of two second-rank tensors. Let $(n)$ be an $n$-dimensional irreducible representation of $\mathit{SO}(3)$. A symmetric second-rank tensor without trace transforms under $(5)$. A trace is a rotation scalar and transforms with respect to the trivial representation (1).

A second-rank tensor with trace transforms under $(1)\oplus (5)$ if it
is symmetric. Without any symmetries present, it is governed
by $(1)\oplus (3)\oplus (5)$. For brevity, we
introduce the short-hand notation $\text{E}:=(1)\oplus (5)$, $\text{B}:=(1)
\oplus (5)$, and $\text{G}:=(3)\oplus (5)$. By working out these products
according to the decomposition rules known from coupling angular
momenta in quantum mechanics, we obtain explicitly:
\begin{subequations}
\label{eq:irreps-so3}
\begin{align}
\label{eq:irreps-EE}
\text{E}\otimes \text{E}&=(1) \oplus (1)' \oplus (5) \oplus (5)' \oplus (9)\,, \displaybreak[0]\\[2ex]
\label{eq:irreps-BB}
\text{B}\otimes \text{B}&=(1) \oplus (1)' \oplus (5) \oplus (5)' \oplus (9)\,, \displaybreak[0]\\[2ex]
\text{G}\otimes \text{G}&=(1) \oplus (5) \oplus (9)\oplus (1)' \oplus (5)' \notag \\
&\phantom{{}={}}\oplus (3)'' \oplus (5)'' \oplus  (7)''\,,
\end{align}
and for the mixed sectors,
\begin{align}
\text{E}\otimes \text{B} &=(1)\oplus (1)'\oplus (5)\oplus (5)' \notag \\
&\phantom{{}={}}\oplus (5)'' \oplus (7)\oplus (9)\,, \displaybreak[0]\\[2ex]
\text{E}\otimes \text{G} &=(3) \oplus (5) \oplus (3)' \oplus (5)'\oplus (7) \notag \\
&\phantom{{}={}}\oplus (1)\oplus (3)'' \oplus (5)'' \oplus (7)' \oplus (9)\,, \displaybreak[0]\\[2ex]
\text{B}\otimes \text{G} &= (3) \oplus (5) \oplus (3)' \oplus (5)'\oplus (7) \notag \\
&\phantom{{}={}}\oplus (1)\oplus (3)'' \oplus (5)'' \oplus (7)' \oplus (9)\,,
\end{align}
\end{subequations}
where primes distinguish different representations of the same dimension from each other.

Interestingly, for each decomposition, there is at least a single rotation scalar transforming under~(1). Such scalars can be treated conveniently in phenomenological analyses, as we shall see below. Appendix~\ref{app:isotropic-configurations} shows in more detail how to compute each of these scalars. Note that some of them are associated with standard dispersion relations.
\begin{table}
\begin{tabular}{cccc}
\toprule
Coeffs. & Definition & \multicolumn{2}{c}{Numbers} \\
\midrule
$K_{\text{EE}}^{ij;kl}$ & $(k_R)^{0i0j;0k0l}$ & $(6\times7)/2=21$ & 15 \\
$K_{\text{BB}}^{ij;kl}$ & $\frac{1}{16}\epsilon^{imn}\epsilon^{jpq}\epsilon^{krs}\epsilon^{luv}(k_R)^{mnpq;rsuv}$ & $(6\times7)/2=21$ & 15 \\
$K_{\text{GG}}^{ij;kl}$ & $\frac{1}{4}\epsilon^{jmn}\epsilon^{lrs}(k_R)^{0imn;0krs}$ & $(8\times9)/2=36$ & 0 \\
$K_{\text{EB}}^{ij;kl}$ & $\frac{1}{4}\epsilon^{kmn}\epsilon^{lrs}(k_R)^{0i0j;mnrs}$ & $6\times6=36$ & 27 \\
$K_{\text{EG}}^{ij;kl}$ & $\frac{1}{2}\epsilon^{lmn}(k_R)^{0i0j;0kmn}$ & $6\times8=48$ & 24 \\
$K_{\text{BG}}^{ij;kl}$ & $\frac{1}{8}\epsilon^{imn}\epsilon^{jpq}\epsilon^{lrs}(k_R)^{mnpq;0krs}$ & $6\times8=48$ & 24 \\
\midrule
Total & & 210 & 105 \\
\bottomrule
\end{tabular}
\caption{Subsets of controlling coefficients transforming under irreducible representations of $\mathit{SO}(3)$. The first column lists the coefficients and the second their definitions. The third column provides the numbers of independent components coefficients from mere index symmetries. The fourth column states the remaining numbers of components after taking into account the second Bianchi identities according to Eq.~\eqref{eq:bianchi-identity-coefficients}.}
\label{tab:subsets-coefficients}
\end{table}
\begin{table}[t]
\begin{tabular}{cccc}
\toprule
Sector & Symmetries & Tracelessness & Reducible rep \\
\midrule
EE & $((i,j),(k,l))$ & No & $[(1)\oplus (5)]\otimes_s [(1)\oplus (5)]$ \\
BB & $((i,j),(k,l))$ & No & $[(1)\oplus (5)]\otimes_s [(1)\oplus (5)]$ \\
GG & $(ij,kl)$ & Both pairs & $[(3)\oplus (5)]\otimes_s [(3)\oplus (5)]$ \\
EB & $(i,j)$ and $(k,l)$ & No & $[(1)\oplus (5)]\otimes [(1)\oplus (5)]$ \\
EG & $(i,j)$ & $2^{\text{nd}}$ pair & $[(1)\oplus (5)]\otimes [(3)\oplus (5)]$ \\
BG & $(i,j)$ & $2^{\text{nd}}$ pair & $[(1)\oplus (5)]\otimes [(3)\oplus (5)]$ \\
\bottomrule
\end{tabular}
\caption{Symmetries and properties of traces for the coefficients $K_{\text{XY}}^{ij;kl}$
with $\text{X,Y}=\text{E, B, G}$ defined in Tab.~\protect\ref{tab:subsets-coefficients}. The GG sector is included, for completeness.
The reason for the tracelessness in $\text{G}$ sectors are relationships for the coefficients imposed from the first Bianchi identities of the Riemann tensor, i.e., we use $\epsilon^{ipq}(k_R)^{oipq\dots}=0$ and similar. The symbol $\otimes_{(s)}$ indicates the (symmetric) direct product of representations.}
\label{tab:properties-coefficients}
\end{table}

In addition to this, we must take into account the second Bianchi identities according to Eqs.~\eqref{eq:EB-next-EB} and \eqref{eq:E-next-B}, since they imply equivalence classes of coefficients. Note that a completely symmetric tensor of rank $r$ with indices that take $n$ possible values has ${n+r-1\choose r}$ independent components. For the EE and BB sectors, there are 6 Bianchi identities each, which reduce the initial 21 components to the remaining 15. Alternatively, according to the generic formula, the number of independent components of a totally symmetric tensor with E-type indices is ${6\choose 4}=15$. Similarly, the BB sector is argued to have 15 remaining coefficients.

For the GG and EB sectors, there are 9 Bianchi identities each, which implies that 27 components are left over from the previous 36. Note also that Eq.~\eqref{eq:EB-identity} relates the GG to the EB coefficients. Hence, it is possible to express one sector
completely in terms of the other. We decided to treat the EB coefficients as the independent ones, i.e., the GG sector is eliminated as an independent set.

Each of the EG and BG sectors has 24 Bianchi identities, which divide the initial 48 components in half. Alternatively, complete symmetry in the E-type or B-type coefficients implies ${5\choose 3}\times 3=30$ coefficients. The 6 Bianchi identities of Eq.~\eqref{eq:EB-identity} then also lead to 24 components. The final numbers of independent component coefficients can be found in the last column of Tab.~\ref{tab:subsets-coefficients}. In fact, the second Bianchi identities eliminate configurations associated with the trivial representation (1) in Eq.~\eqref{eq:irreps-so3}, when these provide the standard dispersion relation.

\subsection{Dispersion relations}
\label{sec:dispersion-relations}

The dispersion relation $p_0=\omega=\omega(\mathbf{p})$ between the gravitational-wave frequency $\omega$ and the spatial wave vector $\mathbf{p}$ must be satisfied, such that the wave equation~\eqref{eq:linearized-field-equations-lorentz-gauge} possesses nontrivial solutions for the metric perturbation. When component coefficients are contracted with $p_0$ components, spurious dispersion relations emerge that are nonperturbative in the SME coefficients~\cite{Schreck:2013kja,Casana:2018rhg,Ferreira:2019lpu}. These are uninteresting within effective field theory and are discarded. Moreover, modes with the standard dispersion relation $\omega(\mathbf{p})=|\mathbf{p}|$ in terms of the spatial wave vector $\mathbf{p}$ are also identified. For a single coefficient, they are interpreted as unphysical gauge modes and will be disregarded, too.

The remaining perturbatively modified dispersion relations are those of the physical modes. For a single generic component coefficient $k$, they are of the form
\begin{equation}
\label{eq:generic-dispersions}
\omega^{(\pm)}(\mathbf{p})\approx |\mathbf{p}| + \Big[f(\mathbf{p})\pm \sqrt{g(\mathbf{p})}\,\Big]k\,,
\end{equation}
with functions $f,g$ depending on $\mathbf{p}$. Recall that Lorentz violation in electromagnetic waves may imply vacuum birefringence, i.e., depending on the polarization of the wave, there are two distinct dispersion relations; see, e.g., Refs.~\cite{Kostelecky:2002hh,Schreck:2026kwd}. A similar phenomenon can occur when spacetime symmetries are violated in linearized gravity. A gravitational wave then splits into two modes with different dispersion relations and propagation velocities. This effect, which Eq.~\eqref{eq:generic-dispersions} reveals, is also known as birefringence in the literature, but now it refers to signal propagation in gravity.

\subsubsection{Computation}

The dispersion relations straightforwardly follow from evaluating the determinant of the $(10\times 10)$ matrix $\bar{M}$ in Eq.~\eqref{eq:free-field-equations-reduced}, setting the expression equal to 0, and solving for $p_0$. Note that $\det(\bar{M})=0$ is not an identity since the gauge has been fixed. Therefore, the dispersion relations can actually be computed in this way for each controlling coefficient. Having these results at hand, they can be covariantized.

There is a second possibility of computing the dispersion relations that is more sophisticated but also more challenging to realize. It was originally conceived for the electromagnetic sector of the SME in Ref.~\cite{Kostelecky:2009zp} and generalized to the gravitational SME at the linearized level in Ref.~\cite{Kostelecky:2017zob}. Its great advantage compared to the straightforward technique is that it does not require gauge fixing and maintains covariance. Let us first lay out the arguments for electromagnetism. The field equations then have the form $M_{\mu}^{\phantom{\mu}\nu}A_{\nu}=0$ with a $(4\times 4)$ matrix $M=(M_{\mu}^{\phantom{\mu}\nu})$ and the four-potential with components $A_{\mu}$. We interpret the matrix $M$ as a linear map $(0,1)\rightarrow (0,1)$ between covariant four-vectors such that $A'_{\mu}=M_{\mu}^{\phantom{\mu}\nu}A_{\nu}$.

Exterior algebra provides a powerful tool set to understand the properties of $M$, amongst them the wedge product denoted as $\wedge$. Consider four linearly independent four-vectors $\{A,B,C,D\}$ such that $A\wedge B\wedge C\wedge D\neq 0$. The linear maps $M$ can be combined to form a multilinear map from the wedge product of initial four-vectors to the wedge product of transformed four-vectors $\{A',B',C',D'\}$:
\begin{subequations}
\begin{align}
A'\wedge B'\wedge C'\wedge D'&=(MA)\wedge (MB)\wedge (MC)\wedge (MD) \notag \\
&=(M\wedge M\wedge M\wedge M)A\wedge B\wedge C\wedge D \notag \\
&:=(\wedge^{(4)}M)A\wedge B\wedge C\wedge D\,,
\end{align}
or in components
\begin{align}
A'_{\mu}\wedge B'_{\nu}\wedge C'_{\rho}\wedge D'_{\sigma}&=(M_{\mu}^{\phantom{\mu}\alpha}\wedge M_{\nu}^{\phantom{\mu}\beta}\wedge M_{\rho}^{\phantom{\rho}\gamma}\wedge M_{\sigma}^{\phantom{\sigma}\delta}) \notag \\
&\phantom{{}={}}\times A_{\alpha}\wedge B_{\beta}\wedge C_{\gamma}\wedge D_{\delta}\,.
\end{align}
\end{subequations}
Here and in the following, $\wedge^{(n)}M$ denotes the $n$-fold wedge product of $M$.

The values of $A'\wedge B'\wedge C'\wedge D'$ and $\wedge^{(4)}M$, respectively, contain information on the rank $r$ of $M$, i.e., the number of linearly independent lines and columns of the linear map $M$. \textit{U}(1) gauge invariance implies that the field equations $M\cdot A=0$ are invariant under $A_{\mu}\rightarrow A_{\mu}+p_{\mu}\chi$ with a scalar function $\chi$. This mandates $M\cdot p=0$, which identifies the gauge modes $A_{\mu}\sim p_{\mu}$. Since $\det(M)\equiv 0$ due to gauge invariance, $\wedge^{(4)}M\equiv 0$ such that $r<4$.

In fact, the solution space of gauge modes is two-dimensional and spanned by a purely timelike and a purely spacelike, longitudinal basis vector. The gauge solutions can be eliminated by one gauge fixing condition and throwing away a single mode obeying $p^2=0$. Thus, the rank of $M$ is $r=2$, whereupon $\wedge^{(3)}M=0$ and $\wedge^{(2)}M\neq 0$. The system $M\cdot A=0$ then exhibits two separate one-dimensional solution spaces $A^{(1)}=\xi\varepsilon^{(1)}$ and $A^{(2)}=\psi\varepsilon^{(2)}$ related to two distinct frequencies $\omega^{(1)}\neq\omega^{(2)}$. Here, $\varepsilon^{(1,2)}$ are transverse basis vectors and $\xi,\psi$ complex parameters.

In fact, $r=1$ is possible, which means that $\wedge^{(2)}M=0$. The system then has a two-dimensional solution space $A=\xi\varepsilon^{(1)}+\psi\varepsilon^{(2)}$, revealing a degeneracy in the frequency, $\omega^{(1)}=\omega^{(2)}=\omega$. This setting includes Maxwell electrodynamics. Thus, the generic solution satisfies $\wedge^{(3)}M=0$. The latter corresponds to the electromagnetic dispersion relation in tensorial form, which has $10^3$ independent components. They are either 0 or involve the scalar dispersion equation multiplied by unimportant factors.

The equation $\wedge^{(3)}M=0$ has 6 free indices. To obtain the scalar dispersion equation, one can follow two approaches. The first benefits from constructing an appropriate Hodge dual, which is a scalar. This method is valuable when one is interested in a covariant form of the scalar dispersion equation. It leads to Eq.~(30) in Ref.~\cite{Kostelecky:2009zp}. The second is a brute-force approach that selects a single nonzero component of $\wedge^{(3)}M=0$ containing the scalar dispersion equation. The latter technique is fruitful when examining specific configurations of SME coefficients.

The arguments are analogous for linearized gravity, but their implementation is technically more involved. Let us generalize the chain of reasoning for electromagnetism to an arbitrary gauge theory with the field equation $M\cdot G=0$ for the $N$-component gauge field $G$. The tensorial dispersion relation then reads $\wedge^{(N-N_g)}M=0$, where $N_g$ is the number of gauge group generators. For electromagnetism, $N=4$ and $N_g=1$.

The field equations of a linearized-gravity theory are of the form $M_{\mu\nu}^{\phantom{\mu\nu}\rho\sigma}h_{\rho\sigma}=0$ with a fourth-rank tensor $M_{\mu\nu}^{\phantom{\mu\nu}\rho\sigma}$ and the metric perturbation with components $h_{\rho\sigma}$. In our case,
\begin{subequations}
\label{eq:linearized-field-equations-h}
\begin{equation}
M_{\mu\nu}^{\phantom{\mu\nu}\rho\sigma}=\tilde{L}_{\mu\nu}^{\phantom{\mu\nu}\rho\sigma}+(\tilde{Q}_R)_{\mu\nu}^{\phantom{\mu\nu}\rho\sigma}\,,
\end{equation}
with the Lichnerowicz operator in momentum space
\begin{align}
\tilde{L}_{\mu\nu}^{\phantom{\mu\nu}\rho\sigma}&=\frac{p^2}{2}(\delta_{\mu}^{\phantom{\mu}\rho}\delta_{\nu}^{\phantom{\nu}\sigma}+\delta_{\mu}^{\phantom{\mu}\sigma}\delta_{\nu}^{\phantom{\nu}\sigma}) \notag \\
&\phantom{{}={}}+\frac{1}{2}\Big[p^{\rho}(p_{\mu}\delta_{\nu}^{\phantom{\nu}\sigma}+p_{\nu}\delta_{\mu}^{\phantom{\mu}\sigma})+p^{\sigma}(p_{\mu}\delta_{\nu}^{\phantom{\nu}\rho}+p_{\nu}\delta_{\mu}^{\phantom{\mu}\rho})\Big] \notag \\
&\phantom{{}={}}+p_{\mu}p_{\nu}\eta^{\rho\sigma}+\eta_{\mu\nu}p^{\rho}p^{\sigma}-p^2\eta_{\mu\nu}\eta^{\rho\sigma}\,,
\end{align}
which has been symmetrized explicitly, and the modification
\begin{equation}
(\tilde{Q}_R)_{\mu\nu}^{\phantom{\mu\nu}\rho\sigma}=-16 (k_R)_{\mu\phantom{\alpha}\nu}^{\phantom{\mu}\alpha\phantom{\nu}\beta\rho\gamma\sigma\delta}p_{\alpha}p_{\beta} p_{\gamma}p_{\delta}\,.
\end{equation}
\end{subequations}
As we explained around Eq.~\eqref{eq:free-field-equations-reduced}, the latter can be written in the form $M\cdot h=0$ or, in components, $M_A^{\phantom{A}B}h_B=0$ with a $(10\times 10)$ matrix $M=(M_A^{\phantom{A}B})$ understood as a linear map and a 10-component vector $h_B$.

Like electromagnetism, linearized gravity possesses a gauge symmetry. The field equations are invariant under linearized diffeomorphisms of the form $h_{\mu\nu}\rightarrow h_{\mu\nu}+p_{\mu}\chi_{\nu}+p_{\nu}\chi_{\mu}$ with a set of four functions $\chi_{\mu}$. Gauge invariance dictates $\det(M)\equiv 0$, which reveals $\wedge^{(10)}M\equiv 0$ and $r<10$.

Since there are four generators of linearized diffeomorphisms, the linear map $M$ must satisfy $\wedge^{(6)}M=0$ and $\wedge^{(5)}M\neq 0$. After all, the metric perturbation $h$ has $N=10$ independent components and $N_g=4$. Hence, the rank of $M$ is $r=5$. When resorting to gauge fixing, the gauge solutions are eliminated by 4 gauge fixing conditions. The remaining 6 components of $h$ are physical, but 4 are auxiliary and do not correspond to propagating modes. An additional requirement is usually implemented to remove one of these 6 auxiliary components, such as the tracelessness of the metric perturbation: $h^{\mu}_{\phantom{\mu}\mu}=0$.

Again, there are two techniques to construct the scalar dispersion equation from $\wedge^{(6)}M=0$. The first requires that a suitable dual be constructed and results in the covariant dispersion equations for gravitational waves given in Eqs.~(14), (15) and (26) of Ref.~\cite{Kostelecky:2017zob}. The second method computes the individual components of $\wedge^{(6)}M=0$. However, doing so is far more challenging than for electromagnetism due to the excessive number of independent components. For a symmetric $M$, the number of independent components of the tensorial dispersion equation $\wedge^{(6)}M=0$ amounts to $55^6$, which is way too large to be evaluated by computer algebra on a personal computer.

Fortunately, we do not need to do so, as the components are either 0 or correspond to the scalar dispersion equation multiplied by an unessential prefactor. So, what we have to do is search for a single nonzero component of $\wedge^{(6)}M$, which directly leads us to the scalar dispersion equation. In practice, this can be achieved via a loop and a random number generator for the components of $\wedge^{(6)}M$, which provides the scalar dispersion equation within mere minutes. The straightforward method based on the equation $\det(\bar{M})=0$ with $\bar{M}$ of Eq.~\eqref{eq:free-field-equations-reduced} and the manifestly covariant one resting upon $\wedge^{(6)}M=0$ with $M$ extracted from Eq.~\eqref{eq:linearized-field-equations-h} were checked to arrive at the same results.

\subsubsection{Explicit findings}

As discussed in Sec.~\ref{sec:number-coefficients}, the second Bianchi identities reduce the 210 independent coefficients deduced from index symmetries to the set of 105 components transforming under the first representation on the right-hand side of Eq.~\eqref{eq:decomposition-product-representation}. Consequently, $\tilde{K}^{\mu\circ\nu\circ\rho\circ\sigma\circ}$ is rendered totally symmetric. We then obtain the following dispersion relations, where $\hat{K}^{\mu\nu\rho\sigma}:=16\tilde{K}^{\mu\circ\nu\circ\rho\circ\sigma\circ}$ is introduced for convenience; see also Eq.~\eqref{eq:definition-K-tilde}. First, we consider coefficients that are not contracted with additional time derivatives in the original field equations~\eqref{eq:linearized-field-equations-lorentz-gauge} in position space. For example, this is the case for the BB sector in Tab.~\ref{tab:subsets-coefficients}. Then, the dispersion relations can be computed at all orders in Lorentz violation. Some coefficients do not lead to vacuum birefringence, such that the dispersion relation for these configurations reads
\begin{equation}
\label{eq:nonbirefringence-BB}
\omega=\sqrt{\mathbf{p}^2-\frac{1}{2}\hat{K}^{\mu\nu}_{\phantom{\mu\nu}\mu\nu}}\,.
\end{equation}
On the contrary, birefringent coefficients imply a more complicated result:
\begin{subequations}
\label{eq:birefringence-BB}
\begin{equation}
\omega^{(\pm)}=\sqrt{\mathbf{p}^2-\frac{1}{4}\hat{K}^{\mu\nu}_{\phantom{\mu\nu}\mu\nu}\pm\sqrt{\Upsilon}}\,,
\end{equation}
with
\begin{align}
\label{eq:definition-Upsilon}
\Upsilon&=\frac{1}{16}\left(\hat{K}^{\mu\nu}_{\phantom{\mu\nu}\mu\nu}\right)^2 \notag \\
&\phantom{{}={}}+\frac{1}{2}\Big(\hat{K}^{\mu\nu\rho\sigma}\hat{K}_{\mu\nu\rho\sigma}-\hat{K}^{\mu\rho\nu}_{\phantom{\mu\rho\nu}\rho}\hat{K}_{\mu\sigma\nu}^{\phantom{\mu\sigma\nu}\sigma}\Big)\,.
\end{align}
\end{subequations}
Note that Eq.~\eqref{eq:nonbirefringence-BB} follows from Eq.~\eqref{eq:birefringence-BB} via the condition
\begin{equation}
\label{eq:birefringence-condition}
\hat{K}^{\mu\nu\rho\sigma}\hat{K}_{\mu\nu\rho\sigma}-\hat{K}_{\mu\rho\nu}^{\phantom{\mu\rho\nu}\rho}\hat{K}^{\mu\sigma\nu}_{\phantom{\mu\sigma\nu}\sigma}=0\,.
\end{equation}
Then the first dispersion relation of Eq.~\eqref{eq:birefringence-BB} is standard, $p_0=|\mathbf{p}|$, whereas the second corresponds to Eq.~\eqref{eq:nonbirefringence-BB}. Thus, Eq.~\eqref{eq:birefringence-condition} can be interpreted as one possible condition for a nonbirefringent vacuum for propagating gravitational waves. We emphasize again that Eqs.~\eqref{eq:nonbirefringence-BB}, \eqref{eq:birefringence-BB} are exact but only hold for configurations such as BB that are not contracted with additional $p_0$ components.

For coefficients contracted with additional time derivatives, spurious modes emerge that are nonperturbative in the SME coefficients and indicate Planck scale effects~\cite{Schreck:2013kja,Casana:2018rhg,Ferreira:2019lpu}. The latter are supposed to be unimportant for phenomenological studies at low frequencies and will be discarded. For such sectors, the dispersion relations obtained are valid at first order in the controlling coefficients. Without birefringence, it holds that
\begin{equation}
\label{eq:nonbirefringence-EE}
\omega \approx |\mathbf{p}|\left(1-\frac{1}{4\mathbf{p}^2}\hat{K}^{\mu\nu}_{\phantom{\mu\nu}\mu\nu}\right)_{p_0=|\mathbf{p}|}\,.
\end{equation}
When birefringence is present,
\begin{equation}
\label{eq:birefringence-EE}
\omega^{(\pm)} \approx |\mathbf{p}|\bigg(1-\frac{1}{8\mathbf{p}^2}\hat{K}^{\mu\nu}_{\phantom{\mu\nu}\mu\nu}\pm \frac{1}{2\mathbf{p}^2}\sqrt{\Upsilon}\,\,\bigg)_{p_0=|\mathbf{p}|}\,,
\end{equation}
with $\Upsilon$ given by Eq.~\eqref{eq:definition-Upsilon}. Equation~\eqref{eq:birefringence-EE} corresponds to Eqs.~(5), (6) of Ref.~\cite{Kostelecky:2016kfm}. However, as will be explained below, our studies go beyond the previous outcomes.

\subsubsection{Lack of total symmetry}
\label{sec:dispersion-without-total-symmetry}

As described within Sec.~\ref{eq:differential-bianchi-identities} in detail, the second Bianchi identities mandate that $k_R$ be completely symmetric. However, it was also explained how the presence of spacetime-dependent scalar fields obstructs that the Bianchi identities are applied. Suitable integrations by parts remove derivatives and generate vector-valued background fields from the scalar fields. Then the results of Eqs.~\eqref{eq:nonbirefringence-BB} -- \eqref{eq:birefringence-EE} do not hold due to the mixing of components with different symmetries. Hence, the dispersion relations for coefficient sets that are not totally symmetric will be determined as follows. They will turn out to be valuable for certain cases.

As before, let us first consider coefficients that are not contracted with additional time derivatives in position space. For the nonbirefringent case,
\begin{equation}
\label{eq:nonbirefringence-new}
\omega=\sqrt{\mathbf{p}^2+\frac{1}{2}(\hat{K}^{\mu\phantom{\mu}\nu}_{\phantom{\mu}\mu\phantom{\nu}\nu}-2\hat{K}^{\mu\nu}_{\phantom{\mu\nu}\mu\nu})}\,.
\end{equation}
Note that for a totally symmetric background, $\hat{K}^{\mu\phantom{\mu}\nu}_{\phantom{\mu}\mu\phantom{\nu}\nu}=\hat{K}^{\mu\nu}_{\phantom{\mu\nu}\mu\nu}$, which reproduces Eq.~\eqref{eq:nonbirefringence-BB}. The birefringent sector is described by
\begin{subequations}
\label{eq:birefringence-new}
\begin{equation}
\omega^{(\pm)}=\sqrt{\mathbf{p}^2+\frac{1}{4}(\hat{K}^{\mu\phantom{\mu}\nu}_{\phantom{\mu}\mu\phantom{\nu}\nu}-2\hat{K}^{\mu\nu}_{\phantom{\mu\nu}\mu\nu})\pm\sqrt{\Xi}}\,,
\end{equation}
with
\begin{align}
\Xi&=\frac{1}{4}\left(\hat{K}^{\mu\nu}_{\phantom{\mu\nu}\mu\nu}\hat{K}^{\rho\phantom{\rho}\sigma}_{\phantom{\rho}\rho\phantom{\sigma}\sigma}-\frac{3}{4}\hat{K}^{\mu\phantom{\mu}\nu}_{\phantom{\mu}\mu\phantom{\nu}\nu}\hat{K}^{\rho\phantom{\rho}\sigma}_{\phantom{\rho}\rho\phantom{\sigma}\sigma}\right) \notag \\
&\phantom{{}={}}+\frac{1}{2}\left(\hat{K}^{\mu\nu\rho\sigma}\hat{K}_{\mu\nu\rho\sigma}-\hat{K}^{\mu\rho\nu}_{\phantom{\mu\rho\nu}\rho}\hat{K}_{\mu\sigma\nu}^{\phantom{\mu\sigma\nu}\sigma}\right)\,.
\end{align}
\end{subequations}
The nonbirefringent dispersion relation~\eqref{eq:nonbirefringence-new} follows from the birefringent one of Eq.~\eqref{eq:birefringence-new} under the condition
\begin{equation}
\frac{1}{16}(\hat{K}^{\mu\phantom{\mu}\nu}_{\phantom{\mu}\mu\phantom{\nu}\nu}-2\hat{K}^{\mu\nu}_{\phantom{\mu\nu}\mu\nu})^2=\Xi\,.
\end{equation}
Then, one of the dispersion relations of Eq.~\eqref{eq:birefringence-new} is standard, whereas the second corresponds to Eq.~\eqref{eq:nonbirefringence-new}. Moreover, a totally symmetric background leads directly to Eq.~\eqref{eq:birefringence-BB}.

For coefficients contracted with additional time derivatives, the results are obtained at first order in the symmetry-violating coefficients. When birefringence is absent,
\begin{equation}
\label{eq:dispersion-relation-nonsymmetric-nonbirefringent}
\omega\approx |\mathbf{p}|\bigg[1+\frac{1}{4\mathbf{p}^2}(\hat{K}^{\mu\phantom{\mu}\nu}_{\phantom{\mu}\mu\phantom{\nu}\nu}-2\hat{K}^{\mu\nu}_{\phantom{\mu\nu}\mu\nu})\bigg]_{p_0=|\mathbf{p}|}\,,
\end{equation}
whereas in the presence of birefringence,
\begin{equation}
\label{eq:dispersion-relation-nonsymmetric-birefringent}
\omega^{(\pm)}\approx |\mathbf{p}|\bigg[1+\frac{1}{8\mathbf{p}^2}(\hat{K}^{\mu\phantom{\mu}\nu}_{\phantom{\mu}\mu\phantom{\nu}\nu}-2\hat{K}^{\mu\nu}_{\phantom{\mu\nu}\mu\nu})\pm \frac{1}{2\mathbf{p}^2}\sqrt{\Xi}\,\bigg]_{p_0=|\mathbf{p}|}.
\end{equation}
For a totally symmetric background, Eqs.~\eqref{eq:dispersion-relation-nonsymmetric-nonbirefringent} and \eqref{eq:dispersion-relation-nonsymmetric-birefringent} reduce to Eqs.~\eqref{eq:nonbirefringence-EE} and \eqref{eq:birefringence-EE}, respectively. Below, we will look into one scenario where these findings are applicable.

\subsubsection{Isotropic modified propagation}
\label{sec:isotropic-dispersion}
While isotropic sets of symmetry-violating coefficients are tied to a specific observer frame only, they play an exceptional role in any study. According to the decompositions of Eq.~\eqref{eq:irreps-so3}, each sector has at least one isotropic coefficient. Suitable traces over the spatial indices of the coefficients in Tab.~\ref{tab:subsets-coefficients} provide rotation scalars; see App.~\ref{app:isotropic-configurations}. In the set of 105 independent coefficients, there are eight isotropic ones, which we define as follows:
\begin{subequations}
\label{eq:isotropic-configurations}
\begin{align}
(\ring{k}_R)_1&:=\frac{1}{3}\sum_{i,j} K_{\mathrm{EE}}^{ii;jj}=\frac{1}{3}\sum_{i,j} (k_R)^{0i0i0j0j}\,, \displaybreak[0]\\[1ex]
(\ring{k}_R)_2&:=\frac{1}{5}\sum_{i,j} \left(k_{\mathrm{EE}}^{ij;ij}-\frac{1}{3}k_{\mathrm{EE}}^{ii;jj}\right) \notag \\
&=\frac{1}{5}\sum_{i,j}\left((k_R)^{0i0j0i0j}-\frac{1}{3}(k_R)^{0i0i0j0j}\right)\,, \displaybreak[0]\\[1ex]
(\ring{k}_R)_3&:=\frac{1}{3}\sum_{i,j} k_{\mathrm{BB}}^{ii;jj}=\frac{1}{12}\sum_{i\dots l}(k_R)^{ijijklkl}\,, \displaybreak[0]\\[1ex]
(\ring{k}_R)_4&:=\frac{1}{5}\sum_{i,j} \left(k_{\mathrm{BB}}^{ij;ij}-\frac{1}{3}k_{\mathrm{BB}}^{ii;jj}\right) \notag \\
&=\frac{1}{20}\sum_{i\dots l}\left((k_R)^{ijklijkl}-\frac{1}{3}(k_R)^{ijijklkl}\right)\,, \displaybreak[0]\\[1ex]
(\ring{k}_R)_5&=\frac{1}{3}\sum_{i,j}K_{\mathrm{EB}}^{ii;jj}=\frac{1}{6}\sum_{i,j,k} (k_R)^{0i0ijkjk}\,, \displaybreak[0]\\[1ex]
(\ring{k}_R)_6&=\frac{1}{2}\sum_iK_{\mathrm{EB}}^{ii;ii} \notag \\
&=\frac{1}{5}\bigg[\frac{1}{3}\sum_{i,j,k}(k_R)^{0i0ijkjk}-\sum_{i,j,k} (k_R)^{0i0jkikj}\bigg]\,, \displaybreak[0]\\[1ex]
\label{eq:definition-k9}
(\ring{k}_R)_7&=\frac{1}{5}\sum_{i,j}K_{\mathrm{EG}}^{ij;ij}=\frac{1}{10}\sum_{\substack{i,p \\ q,r}} \epsilon^{pqr}(k_R)^{0i0p0iqr}\,, \displaybreak[0]\\[1ex]
\label{eq:definition-k10}
(\ring{k}_R)_8&=\frac{1}{5}\sum_{i,j}K_{\mathrm{BG}}^{ij;ij}=\frac{1}{20}\sum_{\substack{i,j,k \\ m,n}} \epsilon^{kmn}(k_R)^{mnij0kij}\,.
\end{align}
\end{subequations}
Indeed, only three of these eight sectors imply modified dispersion relations. At first order in each isotropic coefficient, they are of the form
\begin{subequations}
\label{eq:dispersions-isotropic}
\begin{align}
\label{eq:dispersion-isotropic-EE}
\ring{\omega}_{\mathrm{EE}}&\approx|\mathbf{p}|\left[1-\frac{16}{5}(\ring{k}_R)_1\mathbf{p}^2\right]\,, \displaybreak[0]\\[1ex]
\label{eq:dispersion-isotropic-BB}
\ring{\omega}_{\mathrm{BB}}&=|\mathbf{p}|\left[1-\frac{16}{5}(\ring{k}_R)_3\mathbf{p}^2\right]\,, \displaybreak[0]\\[1ex]
\ring{\omega}_{\mathrm{EB}}&\approx|\mathbf{p}|\left[1-\frac{64}{5}(\ring{k}_R)_5\mathbf{p}^2\right]\,.
\end{align}
\end{subequations}
The EE, BB, and EB sectors possess a single modified isotropic dispersion relation, whereas the EG and BG sectors are devoid of any. The remaining isotropic sectors, whose presence Eq.~\eqref{eq:irreps-so3} indicates, exhibit standard dispersion relations. The second Bianchi identities, as described in Sec.~\ref{eq:differential-bianchi-identities}, either makes them contribute to Eq.~\eqref{eq:dispersions-isotropic} by mixing or gets rid of them.

Note also the similarity between Eqs.~\eqref{eq:dispersion-isotropic-EE} and \eqref{eq:dispersion-isotropic-BB} for the EE and BB sectors, respectively. Furthermore, the definitions of $(\ring{k}_R)_7$ and $(\ring{k}_R)_8$ in Eq.~\eqref{eq:definition-k9} and \eqref{eq:definition-k10}, respectively, are mirror images of each other. These characteristics are courtesy to a duality between the EE and BB, as well as the EG and BG sectors, which is the analog of the well-known dual symmetry of Maxwell electrodynamics.

\subsection{Gravitational Chern-Simons term}

Interestingly, the gravitational CS term poses a significant caveat to implementing the second Bianchi identities according to Sec.~\ref{eq:differential-bianchi-identities}. The fourth Young diagram on the right-hand side of Eq.~\eqref{eq:decomposition-product-representation} corresponds to the irreducible $\mathit{SO}(1,3)$ representation under which $k_R$ transforms when it is chosen according to the EGB term. Therefore, we conclude that the $k_R$ configuration of the CS term must be part of the set that transforms under the irreducible representation of the third Young diagram. The implementation of the Bianchi identities renders $k_R$ totally symmetric in pairs of indices, which is incompatible with the symmetries of the latter.

Resorting to the set of 210 independent coefficients from index symmetries, CS theory of Eq.~\eqref{eq:CS-theory} is reproduced from the isotropic EG and BG sectors of the $k_R$ term with the choices $(\mathring{k}_R)_7=\theta/32=-(\mathring{k}_R)_8$, based on Eqs.~\eqref{eq:definition-k9}, \eqref{eq:definition-k10}; see Tab.~\ref{tab:coefficients-CS} for the explicit values of the coefficients. The linearized CS term does not impact gravitational waves at all when $\theta$ is taken as a constant. In particular, the dispersion relation is the standard one, $\omega(\mathbf{p})=|\mathbf{p}|$. In fact, Eqs.~\eqref{eq:definition-k9}, \eqref{eq:definition-k10} and their combinations have isotropic sectors with standard dispersion relations. As we argued in Sec.~\ref{eq:differential-bianchi-identities}, the gravitational CS term with constant $\theta$ does not contribute to the field equations. The second Bianchi identities can then be applied to remove this configuration from the linearized theory.

Interesting physics potentially occurs for a spacetime-dependent $\theta(x)$. In this case, we linearize the CS Lagrange density such that
\begin{equation}
\mathcal{L}_{\mathrm{CS}}=\frac{\theta}{4\kappa}\epsilon^{\nu\kappa\sigma\tau}\partial_{\tau}(\partial_{\mu}h^{\lambda}_{\phantom{\lambda}\sigma}-\partial^{\lambda}h_{\mu\sigma})\partial_{\kappa}\partial_{\lambda}h^{\mu}_{\phantom{\mu}\nu}\,.
\end{equation}
Suitable integrations by parts and the choice of the Lorenz gauge lead to
\begin{align}
\mathcal{L}_{\mathrm{CS}}&\simeq \frac{\partial_{\kappa}\partial_{\lambda}\theta}{4\kappa}\epsilon^{\nu\kappa\sigma\tau}\partial_{\tau}(\partial_{\mu}h^{\lambda}_{\phantom{\lambda}\sigma}-\partial^{\lambda}h_{\mu\sigma})h^{\mu}_{\phantom{\mu}\nu} \notag \\
&\phantom{{}={}}-\frac{\partial_{\kappa}\theta}{4\kappa}\epsilon^{\nu\kappa\sigma\tau}\partial_{\tau}\square h_{\mu\sigma}h^{\mu}_{\phantom{\mu}\nu}\,.
\end{align}
This then provides the field equations
\begin{align}
0&=\hat{L}^{\mu\nu}+2(\partial_{\kappa}\partial_{\lambda}\theta)\epsilon^{(\mu\kappa\sigma\tau}\partial_{\tau}(\partial^{\nu)}h^{\lambda}_{\phantom{\lambda}\sigma}-\partial^{\lambda}h^{\nu)}_{\phantom{\nu)}\sigma}) \notag \\
&\phantom{{}={}}-2(\partial_{\kappa}\theta)\epsilon^{(\mu\kappa\sigma\tau}\partial_{\tau}\square h^{\nu)}_{\phantom{\mu)}\sigma}\,,
\end{align}
with $\hat{L}^{\mu\nu}$ of Eq.~\eqref{eq:lichnerowicz-operator}. Let us look closer at the choice $\theta=\theta_0t$ with constant $\theta_0$, such that $\partial_{\mu}\theta$ gives rise to a purely timelike, constant background field $(v_{\mu}):=\theta_0(1,0,0,0)$. Then, $\partial_{\mu}\partial_{\nu}\theta=0$ and the field equations result in
\begin{equation}
\label{eq:field-equations-CS-nonconstant-theta}
0=\hat{L}^{\mu\nu}+2\theta_0\epsilon^{0(\mu\sigma\tau}\partial_{\tau}\square h^{\nu)}_{\phantom{\mu)}\sigma}\,.
\end{equation}
The reformulation of the CS term explicitly requires that $k_R$ be spacetime-dependent, even at the linearized level.
\begin{table}
\centering
\begin{tabular}{ccccccc}
\toprule
 & \multicolumn{3}{c}{EG sector} & \multicolumn{3}{c}{BG sector}  \\
\midrule
Coefficients & $K_{\text{EG}}^{ii;ii}$ & $K_{\text{EG}}^{ii;jj}$ & $K_{\text{EG}}^{ij;ij}$ & $K_{\text{BG}}^{ii;ii}$ & $K_{\text{BG}}^{ii;jj}$ & $K_{\text{BG}}^{ij;ij}$ \\[1ex]
Value $\times\sqrt{-g}$ & $\theta/48$ & $-\theta/96$ & $\theta/64$ & $-\theta/48$ & $\theta/96$ & $-\theta/64$ \\
\bottomrule
\end{tabular}
\caption{Nonzero $k_R$ coefficients contributing to gravitational CS theory of Eq.~\eqref{eq:CS-theory} without Eq.~\eqref{eq:bianchi-identity-coefficients} implemented.}
\label{tab:coefficients-CS}
\end{table}

We must now resort to the previously found dispersion relations of Sec.~\ref{sec:dispersion-without-total-symmetry}, which hold for configurations of $k_R$ that are not totally symmetric in index pairs. As we will see in a moment, they incorporate settings related to the $k_R$ term via integrations by parts. One of the four derivatives in the $k_R$ term is then absorbed into the vector field $v_{\mu}$, leading to the following trick. In the dispersion relation, the corresponding momentum is no longer present and must be replaced by $-\mathrm{i}v_{\mu}$. This procedure implies the correct dispersion relation associated with Eq.~\eqref{eq:field-equations-CS-nonconstant-theta} at first order in Lorentz violation.

By doing so, we reproduce the finding already stated in Ref.~\cite{Jackiw:2003pm} that, even in this case, gravitational waves still propagate at the speed of light. Hence, the dispersion relation is standard. In fact, if both $(\mathring{k}_R)_7$ and $(\mathring{k}_R)_8$ remain unspecified, the dispersion relation will be modified, and the modification depends on the sum $(\mathring{k}_R)_7+(\mathring{k}_R)_8$. For the CS term, the EG sector and its dual, the BG sector, contribute with opposite signs, which is why they cancel. This behavior is interpreted as a consequence of the topological character of the Chern-Pontryagin density. It contains both the Riemann tensor and its dual, such that kinematics remain unaffected.

However, as we are now aware of from the previous discussions, the gravitational CS term with spacetime-dependent $\theta(x)$ is physical. Jackiw and Pi elaborated in Ref.~\cite{Jackiw:2003pm} that this term is responsible for modifications at the level of dynamics, e.g., the radiation intensity. A thorough investigation of such effects is beyond the scope of the present project and will be left for a future point in time.

\section{Phenomenology}\label{sec:phenomenology}
The results gained previously allow us to dedicate ourselves to phenomenology. There are two essential questions to answer. The first is whether experimental data are available that can be employed to search for the predicted effects. The second is how to quantify the presence or absence of any nonstandard phenomena in terms of the SME coefficients considered. In the current section, we are going to answer these questions for the nonbirefringent and birefringent sectors of $k_R$. Our analysis will be mainly based on time-of-flight data for different signals of both gravitational and nongravitational origin.

\subsection{Nonbirefringent coefficients}

The isotropic dispersion relations encountered in Sec.~\ref{sec:isotropic-dispersion} are nonbirefringent. To constrain this type of spacetime symmetry violation in linearized gravity, arrival time differences between modified gravitational-wave signals and standard matter particles can be consulted; see, e.g., Ref.~\cite{Nishizawa:2016kba}. In 2017, different observatories independently noticed two highly interesting astrophysical events. In particular, the Advanced LIGO and Virgo detectors measured the gravitational-wave event GW170817 \cite{LIGOScientific:2017vwq}. The Fermi Gamma-ray Burst Monitor detected the gamma-ray burst GRB 170817A \cite{Goldstein:2017mmi,Savchenko:2017ffs} almost simultaneously. It is highly unlikely that both are independent events, i.e., a binary-neutron star merger is considered their same origin. In fact, the gravitational wave event was detected $\unit[(1.74\pm 0.05)]{s}$ before the arriving photons, which allows us to draw conclusions on the propagation velocity difference between the gravitational wave and photons emitted from the same source.

For wave and particle propagation over astrophysical distances, cosmological effects may not be negligible. We assume that our Universe is described by the $\Lambda$CDM model. Modifications of cosmology at the level of the action of Eq.~\eqref{eq:action}, e.g., a cosmological constant, are disregarded. Since the effects on gravitational-wave propagation considered here are already suppressed at first order in the SME coefficients, additional alterations due to cosmology are expected to be suppressed even further. However, we will consider cosmological effects at the level of how a modified dispersion relation gives rise to different arrival times of species propagating over long distances.

Jacob and Piran showed \cite{Jacob:2008bw} how to take into account cosmological expansion when looking at propagation over large distances. They consider a specific massless dispersion relation modified by an isotropic term involving an arbitrary power of the particle momentum, which leads to a subluminal propagation velocity. Two particles of different energies are assumed to be emitted by an unspecified source. The low-energy particle is taken to propagate with $c$, whereas the high-energy particle is subject to a delay. Under these assumptions, their Eq.~(6) describes the arrival time difference of these two particles at the detector. Since our modified dispersion relation is not necessarily of the same form as that considered in Ref.~\cite{Jacob:2008bw}, their result needs to be generalized.

Let $\Delta v=c\delta v=c\delta v(\mathbf{p})$ be the velocity difference of two particles or wave fronts at a given momentum or wave vector $\mathbf{p}$. Consider a source located at redshift $z$, where the detector ought to be at $z=0$. By following the arguments of Ref.~\cite{Jacob:2008bw}, the arrival time difference $\Delta t$ between the particle of modified velocity $v=c(1+\delta v)$ compared to the particle with standard propagation velocity $v=c$ is generically given by
\begin{subequations}
\label{eq:arrival-time-difference-JP}
\begin{align}
\Delta t&=-\frac{1}{H_0} \int_0^z \mathrm{d}z'\,\frac{\delta v\left[(1+z')\mathbf{p}\right]}{\Xi(z')}\,, \displaybreak[0]\\[2ex]
\Xi(z')&=\Big[\Omega_r(1+z')^4+\Omega_m(1+z')^3 \notag \\
&\phantom{{}={}}\,+\Omega_k(1+z')^2+\Omega_{\Lambda}\Big]^{1/2}\,,
\end{align}
\end{subequations}
where $H_0$ is the present-day Hubble parameter and $\Omega_{r,m,k,\Lambda}$ are the normalized, dimensionless densities related to radiation, matter, curvature, and dark energy. For the cosmological parameters, we employ the experimental values resulting from fits to the $\Lambda$CDM model. Once the parameters and modified dispersion relation are given explicitly, the latter integral can be evaluated numerically. On the one hand, for $\delta v>0$, we have $\Delta t<0$, since the modified particle arrives before the standard one. On the other hand, for $\delta v<0$, it holds that $\Delta t>0$, which means that the modified particle is delayed compared to the standard one.

Let us first look at isotropic, nonbirefringent cases, since these are the simplest ones to consider. The dispersion relations found in Sec.~\ref{sec:isotropic-dispersion} are generically of the form
\begin{equation}
\ring{\omega}=|\mathbf{p}|\Big(1+\zeta\ring{k}_R\mathbf{p}^2\Big)\,,
\end{equation}
with a dimensionless real parameter $\zeta$ and an isotropic coefficient $\ring{k}_R$; cf.~Eq.~\eqref{eq:dispersions-isotropic}. The propagation velocity corresponds to the group velocity, which, for isotropic settings, is given by
\begin{equation}
\label{eq:group-velocity-isotropic}
v=\frac{\partial\ring{\omega}}{\partial|\mathbf{p}|}\,.
\end{equation}
Thus, the velocity difference $\delta v$ compared to a standard photon amounts to
\begin{equation}
\label{eq:velocity-difference}
\delta v=3\zeta\ring{k}_R\mathbf{p}^2\approx 3\zeta\ring{k}_R\frac{\omega^2}{c^2}\,,
\end{equation}
which is correct at first order in $\ring{k}_R$ when the standard dispersion relation $\omega=c|\mathbf{p}|$ is used. The latter result exhibits dispersion, and the modification becomes ever more dominant for rising frequency, since we are dealing with a dim-6 operator. Note that $[\ring{k}_R]=\mathrm{GeV}^{-2}$, or equivalently, $[\ring{k}_R]=\mathrm{m}^2$ in natural units, where our preference is the latter. After all, it is more common in gravity to work with length scales instead of energy scales.

Based on the time delay between the gravitational-wave event GW170817~\cite{LIGOScientific:2017vwq} and the photons from GRB 170817A~\cite{Goldstein:2017mmi,Savchenko:2017ffs}, we can place bounds on dim-6 spacetime symmetry violation in gravitational waves. There are two possibilities for doing so, resting upon different assumptions. In general, any kind of Lorentz violation in the photon sector is disregarded in the present work. In the first scenario, the gravitational wave propagates faster than the photons. After both signals are emitted simultaneously at the source and propagate to the detector on Earth, the photons are delayed by $\Delta T$. In the second scenario, the gravitational wave is slower than the photons, but it is emitted some time $(\Delta T)_{\mathrm{int}}$ before them. Thus, the photons are intrinsically delayed, but they catch up to the gravitational wave, such that the remaining time difference is $(\Delta T)_{\mathrm{int}}-\Delta T$. Intrinsic time delays are expected to depend on the mechanism of the merger of the progenitor objects. Both nuclear physics and gravity simultaneously play a significant role in these processes. However, the details of what exactly happens during a merger are still unknown. Therefore, $(\Delta T)_{\mathrm{int}}$ can only be estimated, and we choose $(\Delta T)_{\mathrm{int}}=\unit[10]{s}$ as a conservative value.

Our concrete point of departure is the isotropic coefficients defined in Eq.~\eqref{eq:irreps-so3}. To constrain them, we use the cosmological parameters $\Omega_r=0=\Omega_k$, $\Omega_m=0.311$, $\Omega_{\Lambda}= 0.689$, and $H_0=\unit[67.7\times 10^3]{m\cdot s^{-1}\cdot Mpc^{-1}}$ according to the Planck results of 2018~\cite{Planck:2018vyg}. Moreover, $\omega=2\pi f$, where $f=\unit[100]{Hz}$ is taken as a conservative estimate of the gravitational-wave frequency. The luminosity distance to the event was measured to be $D=\unit[40]{Mpc}$. This is a cosmologically small distance, whereupon the Hubble law $z=H_0D/c$ provides the approximate redshift. 

The direction $\hat{\mathbf{n}}$ of the source position is usually reported in terms of the declination $\delta$ and right ascension $\alpha$. Throughout this work, we employ the spherical polar angles $(\vartheta,\varphi)$, which are related to the astronomical coordinates through $\vartheta=\pi/2-\delta$ and $\varphi=\alpha$, respectively. Accordingly, the sky location of the source is described as
\begin{equation}
\hat{\mathbf{n}}=\begin{pmatrix}
\sin\vartheta\cos\varphi \\
\sin\vartheta\sin\varphi \\
\cos\vartheta \\
\end{pmatrix}=\begin{pmatrix}
\cos\delta\cos\alpha \\
\cos\delta\sin\alpha \\
\sin\delta \\
\end{pmatrix}\,.
\end{equation}
Explicitly, the declination and right ascension of the event GW170817/GRB170817A are given by $\delta=-23^{\circ}22'53.37''$ and $\alpha=13^{\mathrm{h}}09^{\mathrm{m}}48.08^{\mathrm{s}}$, which correspond to the angles $\vartheta\approx 113.4^{\circ}$ and $\varphi\approx 197.5^{\circ}$, respectively. Note that the propagation direction $\hat{\mathbf{p}}$ of the gravitational-wave event is opposite to the sky location of the source: $\hat{\mathbf{p}}=-\hat{\mathbf{n}}$.

We will perform a maximum-reach analysis, i.e., SME coefficients are taken as nonzero, only one at a time. This technique assumes that accidental cancelations between individual coefficients do not occur, which leads to the strictest constraints on spacetime symmetry violation from experimental data. Equations~\eqref{eq:arrival-time-difference-JP} and \eqref{eq:velocity-difference} in combination with Eq.~\eqref{eq:dispersions-isotropic} then imply the following two-sided bounds on the observable isotropic coefficients:
\begin{subequations}
\label{eq:constraints-isotropic}
\begin{align}
\unit[-9\times 10^{-6}]{m^2}&<(\ring{k}_R)_{1,3}<\unit[4\times 10^{-5}]{m^2}\,, \\[1ex]
\unit[-2\times 10^{-6}]{m^2}&<(\ring{k}_R)_5<\unit[1\times 10^{-5}]{m^2}\,.
\end{align}
\end{subequations}
Next, we turn to the nonbirefringent coefficients of Tab.~\ref{tab:nonbirefringent-coefficients}, which all lead to anisotropies in gravitational-wave propagation. The modified dispersion relation at first order in the SME coefficients is generically of the form
\begin{equation}
\omega\approx |\mathbf{p}|\left(1-(k_R)^X\frac{f(\mathbf{p})}{\mathbf{p}^2}\right)\,,
\end{equation}
where $(k_R)^X$ denotes a single one of the controlling coefficients considered with a set $\{X\}$ of indices, and $f(\mathbf{p})$ is a function that depends on the spatial components $\mathbf{p}$ of the wave four-vector.

For an anisotropic coefficient, the propagation velocity must be obtained from the group velocity three-vector:
\begin{equation}
\label{eq:group-velocity-anisotropic}
v=\left|\frac{\partial\omega}{\partial\mathbf{p}}\right|\,.
\end{equation}
In general, Eq.~\eqref{eq:group-velocity-isotropic} differs from the latter Eq.~\eqref{eq:group-velocity-anisotropic} when signal propagation is anisotropic. Both formulas provide coinciding results only for isotropic cases.
\begin{table}[t]
\begin{tabular}{ccrc}
\toprule
Sector & Coefficient & \multicolumn{1}{c}{$\xi$} & $f(\theta,\phi)$ \\
\midrule
EE & $(k_R)^{01010101}$ & 12 & $(c_{\theta}^2+s_{\theta}^2s_{\phi}^2)^2$ \\[0.4ex]
     & $(k_R)^{02020202}$ & 12 & $(c_{\theta}^2+s_{\theta}^2c_{\phi}^2)^2$ \\[0.4ex]
     & $(k_R)^{03030303}$ & 12 & $s_{\theta}^4$ \\
\midrule
BB & $(k_R)^{12121212}$ & 12 & $s_{\theta}^4$ \\[0.4ex]
      & $(k_R)^{13131313}$ & 12 & $(c_{\theta}^2+s_{\theta}^2c_{\phi}^2)^2$ \\[0.4ex]
      & $(k_R)^{23232323}$ & 12 & $(c_{\theta}^2+s_{\theta}^2s_{\phi}^2)^2$ \\
\bottomrule
\end{tabular}
\caption{Nonbirefringent coefficients among the six different subsets of coefficients of Tab.~\ref{tab:subsets-coefficients}. The first column states the sector and the second the nonbirefringent coefficients identified. The third and fourth columns provide the dimensionless prefactor $\xi$ and the angular-dependent function $f$ of the generic propagation velocity~\eqref{eq:group-velocity-nonbirefringent}. For brevity, $s_x=\sin(x)$ and $c_x=\cos(x)$.}
\label{tab:nonbirefringent-coefficients}
\end{table}

As of now, we will explicitly be using spherical coordinates $(p,\theta,\phi)$ in the frequency domain, where the radial coordinate is $p:=|\mathbf{p}|$ and $\theta$, $\phi$ are the polar and azimuthal angles of the wave vector. The latter are related to the angles of the sky location of the event by $\theta=\pi-\vartheta$ and $\phi=\pi+\varphi$. For an anisotropic, nonbirefringent coefficient, there is a single mode whose propagation velocity exhibits an angular dependence:
\begin{equation}
\label{eq:group-velocity-nonbirefringent}
v\approx 1-\xi (k_R)^Xp^2f(\theta,\phi)\,,
\end{equation}
where $\xi$ is a dimensionless prefactor. Since we are dealing with a dim-6 operator originally, the modification rises with the modulus squared of the spatial wave vector. The angular dependence has been separated from the frequency dependence of the result and is contained in the dimensionless function $f(\theta,\phi)$. In principle, the latter can be expanded in terms of spherical harmonics, as is frequently done to analyze angular behaviors~\cite{Kostelecky:2009zp,Diaz:2013wia,Kostelecky:2016kfm}. This process introduces new coefficients that are bounded based on experimental data. However, we decided to work with the coefficients of the background field $k_R$ in the linearized wave equation~\eqref{eq:linearized-field-equations-lorentz-gauge}.
\begin{table}
\begin{tabular}{cccc}
\toprule
Sector & Lower bound [$\mathrm{m}^2$] & Coefficient & Upper bound [$\mathrm{m^2}$] \\
\midrule
EE & $-1\times 10^{-4}<$ & $(k_R)^{01010101}$ & $<6\times 10^{-4}$ \\
   & $-9\times 10^{-6}<$ & $(k_R)^{02020202}$ & $<4\times 10^{-5}$ \\
   & $-1\times 10^{-5}<$ & $(k_R)^{03030303}$ & $<5\times 10^{-5}$ \\
\midrule
BB & $-1\times 10^{-5}<$ & $(k_R)^{12121212}$ & $<5\times 10^{-5}$ \\
   & $-9\times 10^{-6}<$ & $(k_R)^{13131313}$ & $<4\times 10^{-5}$ \\
   & $-1\times 10^{-4}<$ & $(k_R)^{23232323}$ & $<6\times 10^{-4}$ \\
\bottomrule
\end{tabular}
\caption{Two-sided constraints on the nonbirefringent coefficients from Tab.~\protect\ref{tab:nonbirefringent-coefficients}. The first column states the sectors of Tab.~\ref{tab:subsets-coefficients} and the third column the coefficient considered. The second and fourth columns provide the lower and upper bounds, respectively.}
\label{tab:bounds-nonbirefringent-coefficients}
\end{table}

Explicit results for all nonbirefringent coefficients are stated in Tab.~\ref{tab:nonbirefringent-coefficients}. Several remarks are in order. First, the sign in Eq.~\eqref{eq:group-velocity-nonbirefringent} has been chosen such that $\xi$ is positive. Second, it is interesting to note that each function $f(\theta,\phi)$ is manifestly nonnegative for any choice of $\theta$ and $\phi$. Thus, it is, in fact, the sign of $\xi$ that determines whether the gravitational wave propagates slower or faster than light. For all nonbirefringent coefficients, propagation is subluminal. Third, the same angular behavior $f(\theta,\phi)$ occurs repeatedly for different coefficients. This observation reveals a certain degeneracy that was already mentioned when we decomposed the reducible product of $\mathit{SO}(1,3)$ representations in Sec.~\ref{sec:number-coefficients}. Fourth, note the dualities between different individual coefficients or subsets of coefficients, expressed via the correspondence $\{(01),(02),(03)\}\leftrightarrow \{(23),(13),(23)\}$ between pairs of E- and B-type indices. The behaviors of dual coefficients are the same, which poses an excellent crosscheck of the results. Last but not least, the sectors EB, EG, and BG are devoid of nonbirefringent coefficients.

Now, the method employed previously to constrain the isotropic combinations of coefficients, cf.~Eq.~\eqref{eq:constraints-isotropic}, is adopted to bound the anisotropic, nonbirefringent coefficients of Tab.~\ref{tab:nonbirefringent-coefficients}. The values are given in Tab.~\ref{tab:bounds-nonbirefringent-coefficients}. As before, all bounds are two-sided, but the intervals around zero are asymmetric. This asymmetry is a consequence of the choice of $(\Delta T)_{\mathrm{int}}$ related to our ignorance of intrinsic time delay. In principle, the intervals could be made symmetric by choosing $(\Delta T)_{\mathrm{int}}$ properly. However, such a procedure is artificial and not motivated by any physics that plays a role during the merger. The lower bounds range between $\unit[-1\times 10^{-4}]{m^2}$ and $\unit[-9\times 10^{-6}]{m^2}$, whereas the upper ones lie between $\unit[4\times 10^{-5}]{m^2}$ and $\unit[6\times 10^{-4}]{m^2}$. Thus, the validity of standard linearized gravity is tested down to the millimeter regime.

In the following, we estimate the sensitivity to bound a single generic nonbirefringent coefficient $k$:
\begin{align}
\label{eq:sensitivity-estimate-nonbirefringent}
|k|&\lesssim\unit[10^{-4}]{m^2}\frac{\Delta D}{\unit[500\,000]{km}}\left(\frac{\unit[40]{Mpc}}{D}\right)\left(\frac{\frac{2\pi}{\unit[3000]{km}}}{|\mathbf{p}|}\right)^2 \notag \\
&\approx\unit[10^{-4}]{m^2}\frac{\Delta D}{\unit[5\times 10^8]{m}}\left(\frac{\unit[10^{24}]{m}}{D}\right)\left(\frac{\lambda}{\unit[3\times 10^6]{m}}\right)^2\,,
\end{align}
where $D$ is the distance to the source GW170817/GRB 170817A, $\Delta D$ is the difference between the propagation lengths of photons and the degenerate gravitational-wave modes, and $\lambda$ the wavelength of the gravitational wave.

Sensitivity increases when multimessenger events are detected, whose sources are farther away from Earth, for decreasing arrival time differences measured and for smaller wavelengths. After all, the kinematic method of constraining spacetime symmetry violation used here relies on the accumulation of sensitivity during propagation. Nonminimal coefficients introduce a fundamental length scale into the theory, where physics beyond GR is expected to occur. Gravitational-wave signals can probe this new physics more effectively the smaller their wavelength is.

\subsection{Birefringent coefficients}
The controlling coefficients beyond those listed in Tab.~\ref{tab:nonbirefringent-coefficients} lead to birefringence for gravitational waves. These cases are characterized by two distinct modes with different dispersion relations and propagation velocities. The latter are always anisotropic, and we have not identified a set of birefringent coefficients leading to isotropic propagation; cf.~Eq.~\eqref{eq:dispersions-isotropic}. The dispersion relations and propagation velocities are more involved compared to the nonbirefringent sector, and it would be unreasonable to list each one of these. In general, they are written in the form
\begin{equation}
\label{eq:group-velocity-birefringent}
v^{(\pm)}\approx 1-\xi^{(\pm)}(k_R)^Xp^2f^{(\pm)}(\theta,\phi)\,,
\end{equation}
with dimensionless numbers $\xi^{(\pm)}$, which are not necessarily equal, and two different angular functions $f^{(\pm)}(\theta,\phi)$. Table~\ref{tab:birefringent-coefficients} provides a simple example for each of the five sectors of independent coefficients.
\begin{table}[t]
\begin{tabular}{ccrc}
\toprule
Sector & Coefficient & \multicolumn{1}{c}{$\xi^{(\pm)}$} & $f^{(\pm)}(\theta,\phi)$ \\
\midrule
EE & $(k_R)^{01030303}$ & $-24$ & $s^3_{\theta}\big(c_{\theta}c_{\phi}\pm\sqrt{1-s_{\theta}^2c_{\phi}^2}\big)$ \\[1ex]
BB & $(k_R)^{12121213}$ & 24 & $s^3_{\theta}\big(c_{\theta}s_{\phi}\pm\sqrt{1-s_{\theta}^2s_{\phi}^2}\big)$ \\[1ex]
EB & $(k_R)^{01011212}$ & $-6$ & $s^2_{2\theta}c^2_{\phi}$ \\[0.7ex]
   &                & 24 & $[3-(2-c_{2\theta})c_{\phi}^2]s_{\theta}^2$ \\[0.7ex]
BG & $(k_R)^{01121212}$ & 24 & $s^3_{\theta}\big(s_{\phi}\pm \sqrt{1-s_{\theta}^2c_{\phi}^2}\big)$ \\[0.7ex]
EG & $(k_R)^{01010112}$ & 24 & $s_{\theta}(1-s_{\theta}^2c_{\phi}^2)\big(s_{\phi}\pm\sqrt{1-s_{\theta}^2c_{\phi}^2}\big)$ \\
\bottomrule
\end{tabular}
\caption{Examples for birefringent coefficients. The first column lists the sector of Tab.~\ref{tab:subsets-coefficients} and the second an exemplary coefficient. The third and fourth columns state the dimensionless numbers $\xi^{(\pm)}$ and the angular functions $f^{(\pm)}(\theta,\phi)$, respectively, of Eq.~\eqref{eq:group-velocity-nonbirefringent}.}
\label{tab:birefringent-coefficients}
\end{table}

Several comments are in order. First, the dimensionless parameters $\xi^{(\pm)}$ are again of the form $3\times 2^n$, where $n$ covers a subset of $\mathbb{N}$. Positive and negative values for $\xi^{(\pm)}$ are encountered. Second, the majority of the functions $f^{(\pm)}(\theta,\phi)$ are much more complicated than the analogous ones in nonbirefringent settings; cf.~Tab.~\ref{tab:nonbirefringent-coefficients}. For example, the latter can now involve square root functions. Moreover, for some coefficients, $f^{(\pm)}(\theta,\phi)$ are mapped onto each other by merely interchanging certain signs. However, they may also be substantially different.
\begin{table*}
\begin{tabular}{clclcl}
\toprule
Coefficient & Bound [$\mathrm{m^2}$] & Coefficient & Bound [$\mathrm{m^2}$] & Coefficient & Bound [$\mathrm{m^2}$] \\
\midrule
\multicolumn{2}{c}{EE} & \multicolumn{2}{c}{BB} & \multicolumn{2}{c}{EB} \\
\midrule
$|(k_R)^{01010102}|$ & $<3\times 10^{-10}$ & $|(k_R)^{12121213}|$ & $<8\times 10^{-9}$ & $|(k_R)^{01011212}|$ & $<2\times 10^{-9}$ \\
$|(k_R)^{01010103}|$ & $<9\times 10^{-10}$ & $|(k_R)^{12121223}|$ & $<8\times 10^{-9}$ & $|(k_R)^{01011213}|$ & $<3\times 10^{-10}$ \\
$|(k_R)^{01010202}|$ & $<2\times 10^{-10}$ & $|(k_R)^{12121313}|$ & $<2\times 10^{-9}$ & $|(k_R)^{01011223}|$ & $<3\times 10^{-10}$ \\
$|(k_R)^{01010203}|$ & $<3\times 10^{-10}$ & $|(k_R)^{12121323}|$ & $<9\times 10^{-10}$ & $|(k_R)^{01011313}|$ & $<2\times 10^{-10}$ \\
$|(k_R)^{01010303}|$ & $<2\times 10^{-9}$ & $|(k_R)^{12122323}|$ & $<2\times 10^{-9}$ & $|(k_R)^{01011323}|$ & $<1\times 10^{-10}$ \\
$|(k_R)^{01020202}|$ & $<4\times 10^{-10}$ & $|(k_R)^{12131313}|$ & $<1\times 10^{-9}$ & $|(k_R)^{01012323}|$ & $<2\times 10^{-10}$ \\
$|(k_R)^{01020203}|$ & $<3\times 10^{-10}$ & $|(k_R)^{12131323}|$ & $<3\times 10^{-10}$ & $|(k_R)^{01021212}|$ & $<9\times 10^{-10}$ \\
$|(k_R)^{01020303}|$ & $<9\times 10^{-10}$ & $|(k_R)^{12132323}|$ & $<3\times 10^{-10}$ & $|(k_R)^{01021213}|$ & $<2\times 10^{-10}$ \\
$|(k_R)^{01030303}|$ & $<8\times 10^{-9}$ & $|(k_R)^{12232323}|$ & $<9\times 10^{-10}$ & $|(k_R)^{01021223}|$ & $<2\times 10^{-10}$ \\
$|(k_R)^{02020203}|$ & $<1\times 10^{-9}$ & $|(k_R)^{13131323}|$ & $<4\times 10^{-10}$ & $|(k_R)^{01021313}|$ & $<1\times 10^{-10}$ \\
$|(k_R)^{02020303}|$ & $<2\times 10^{-9}$ & $|(k_R)^{13132323}|$ & $<2\times 10^{-10}$ & $|(k_R)^{01021323}|$ & $<6\times 10^{-11}$ \\
$|(k_R)^{02030303}|$ & $<8\times 10^{-9}$ & $|(k_R)^{13232323}|$ & $<3\times 10^{-10}$ & $|(k_R)^{01022323}|$ & $<1\times 10^{-10}$ \\
\cmidrule{1-4}
\multicolumn{2}{c}{EG}                    & \multicolumn{2}{c}{BG}                     & $|(k_R)^{01031313}|$ & $<3\times 10^{-10}$ \\
\cmidrule{1-4}
$|(k_R)^{01010112}|$ & $<9\times 10^{-10}$ & $|(k_R)^{01121212}|$ & $<8\times 10^{-9}$ & $|(k_R)^{01031323}|$ & $<2\times 10^{-10}$ \\
$|(k_R)^{01010113}|$ & $<3\times 10^{-10}$ & $|(k_R)^{01121213}|$ & $<9\times 10^{-10}$  & $|(k_R)^{01032323}|$ & $<3\times 10^{-10}$ \\
$|(k_R)^{01010123}|$ & $<8\times 10^{-11}$ & $|(k_R)^{01121223}|$ & $<9\times 10^{-10}$ & $|(k_R)^{02021212}|$ & $<2\times 10^{-9}$ \\
$|(k_R)^{01010212}|$ & $<3\times 10^{-10}$ & $|(k_R)^{01121313}|$ & $<3\times 10^{-10}$ & $|(k_R)^{02021213}|$ & $<4\times 10^{-10}$ \\
$|(k_R)^{01010223}|$ & $<6\times 10^{-11}$ & $|(k_R)^{01121323}|$ & $<2\times 10^{-10}$ & $|(k_R)^{02021223}|$ & $<3\times 10^{-10}$ \\
$|(k_R)^{01010312}|$ & $<1\times 10^{-10}$ & $|(k_R)^{01122323}|$ & $<3\times 10^{-10}$ & $|(k_R)^{02021313}|$ & $<3\times 10^{-10}$ \\
$|(k_R)^{01010313}|$ & $<3\times 10^{-10}$ & $|(k_R)^{01131313}|$ & $<4\times 10^{-10}$ & $|(k_R)^{02021323}|$ & $<1\times 10^{-10}$ \\
$|(k_R)^{01010323}|$ & $<3\times 10^{-10}$ & $|(k_R)^{01131323}|$ & $<1\times 10^{-10}$ & $|(k_R)^{02022323}|$ & $<2\times 10^{-10}$ \\
$|(k_R)^{01020212}|$ & $<3\times 10^{-10}$ & $|(k_R)^{01132323}|$ & $<1\times 10^{-10}$ & $|(k_R)^{02031313}|$ & $<4\times 10^{-10}$ \\
$|(k_R)^{01020223}|$ & $<9\times 10^{-11}$ & $|(k_R)^{01232323}|$ & $<3\times 10^{-10}$ & $|(k_R)^{02031323}|$ & $<2\times 10^{-10}$ \\
$|(k_R)^{01030212}|$ & $<1\times 10^{-10}$ & $|(k_R)^{02121212}|$ & $<8\times 10^{-9}$ & $|(k_R)^{02032323}|$ & $<3\times 10^{-10}$ \\
$|(k_R)^{01020313}|$ & $<2\times 10^{-10}$ & $|(k_R)^{02121213}|$ & $<1\times 10^{-9}$ & $|(k_R)^{03031313}|$ & $<2\times 10^{-9}$ \\
$|(k_R)^{01020323}|$ & $<2\times 10^{-10}$ & $|(k_R)^{02121223}|$ & $<9\times 10^{-10}$ & $|(k_R)^{03031323}|$ & $<9\times 10^{-10}$ \\
$|(k_R)^{01030313}|$ & $<9\times 10^{-10}$ & $|(k_R)^{02121313}|$ & $<4\times 10^{-10}$ & $|(k_R)^{03032323}|$ & $<2\times 10^{-9}$ \\
$|(k_R)^{01030323}|$ & $<9\times 10^{-10}$ & $|(k_R)^{02121323}|$ & $<2\times 10^{-10}$ &                      &                     \\
$|(k_R)^{02020212}|$ & $<1\times 10^{-9}$ & $|(k_R)^{02122323}|$ & $<3\times 10^{-10}$ &                      &                     \\
$|(k_R)^{02020223}|$ & $<4\times 10^{-10}$ & $|(k_R)^{02131313}|$ & $<4\times 10^{-10}$ &                      &                     \\
$|(k_R)^{02020312}|$ & $<4\times 10^{-10}$ & $|(k_R)^{02131323}|$ & $<1\times 10^{-10}$ &                      &                     \\
$|(k_R)^{02020313}|$ & $<4\times 10^{-10}$ & $|(k_R)^{02231323}|$ & $<1\times 10^{-10}$ &                      &                     \\
$|(k_R)^{02020323}|$ & $<3\times 10^{-10}$ & $|(k_R)^{02232323}|$ & $<3\times 10^{-10}$ &                     &                     \\
$|(k_R)^{02030313}|$ & $<1\times 10^{-9}$ & $|(k_R)^{03131313}|$ & $<1\times 10^{-9}$ &                      &                     \\
$|(k_R)^{02030323}|$ & $<9\times 10^{-10}$ & $|(k_R)^{03131323}|$ & $<3\times 10^{-10}$ &                      &                     \\
$|(k_R)^{03030313}|$ & $<8\times 10^{-9}$ & $|(k_R)^{03132323}|$ & $<3\times 10^{-10}$ &                      &                     \\
$|(k_R)^{03030323}|$ & $<8\times 10^{-9}$ &  $|(k_R)^{03232323}|$ & $<9\times 10^{-10}$ &                      &                     \\
\bottomrule
\end{tabular}
\caption{Bounds on birefringent coefficients of the EE, BB, EB, EG, and BG sectors.}
\label{tab:constraints-birefringence}
\end{table*}

The birefringent coefficients can be constrained with a single gravitational-wave signal, as long as it is impossible to identify two distinct modes. We intend to base this analysis on the first gravitational-wave event ever observed by LIGO, which is GW150914~\cite{LIGOScientific:2016aoc}. The distance of the latter is $D\approx \unit[410]{Mpc}$, which amounts to the redshift $z\approx 0.09$. Its sky position is given by $\theta\approx 160^{\circ}$ and $\phi\approx 120^{\circ}$. Recall again that the propagation direction $\hat{\mathbf{p}}$ of the event is opposite to its position vector.

The signal ranges over approximately $\unit[0.1]{s}$, and a splitting in the frequency domain is not evident. As done in Ref.~\cite{Kostelecky:2016kfm}, we estimate that if a splitting is present, the arrival time difference $\Delta t$ of the two modes must be smaller than \unit[0.003]{s}. The distinct velocities of Eq.~\eqref{eq:group-velocity-birefringent} amount to a velocity difference between the modes:
\begin{equation}
\delta v:=|v^{(+)}-v^{(-)}|\,.
\end{equation}
Equation~\eqref{eq:arrival-time-difference-JP} then gives the arrival time difference of the two modes at the detector. Since a latter is not observed in the experimental signal of Ref.~\cite{LIGOScientific:2016aoc}, the birefringent coefficients are constrained. The explicit bounds are to be found in Tab.~\ref{tab:constraints-birefringence}. The norms of these constraints range from $\unit[8\times 10^{-9}]{m^2}$ to $\unit[2\times 10^{-10}]{m^2}$, i.e., standard linearized gravity is confirmed to be valid down to 10 microns. Due to anisotropy, sensitivity depends on the sky location of the event consulted. Choosing an alternative event with similar frequency, but at a sky location far from that of GW150914, will change the constraints of Tab.~\ref{tab:constraints-birefringence}, while the overall range is expected remain the same.

We estimate the sensitivity to birefringence in gravitational waves for a generic coefficient $k$ of the dim-6 operator considered as follows:
\begin{align}
\label{eq:sensitivity-estimate-birefringent}
|k|&\lesssim\unit[2\times 10^{-8}]{m^2}\frac{\Delta D}{\unit[900]{km}}\left(\frac{\unit[410]{Mpc}}{D}\right)\left(\frac{\frac{2\pi}{\unit[3000]{km}}}{|\mathbf{p}|}\right)^2 \notag \\
&\approx\unit[2\times 10^{-8}]{m^2}\frac{\Delta D}{\unit[9\times 10^5]{m}}\left(\frac{\unit[10^{25}]{m}}{D}\right)\left(\frac{\lambda}{\unit[3\times 10^6]{m}}\right)^2\,,
\end{align}
where $D$ is the distance to the source of GW150914, $\Delta D$ is the difference between the propagation lengths of both modes, and $\lambda$ the wavelength.

Similarly to Eq.~\eqref{eq:sensitivity-estimate-nonbirefringent} for the nonbirefringent coefficients, sensitivity increases with rising $D$, as well as with decreasing $\Delta D$ and $\lambda$, respectively. The larger sensitivity for birefringent coefficients compared to nonbirefringent ones is a direct implication of the upper limit for the arrival time difference estimated from the signal GW150914. Inspection of the chirp allows for inferring an upper limit that is orders of magnitude smaller than the actual $\Delta T$ for the multimessenger event GW170817/GRB 170817A.

\section{Conclusions and outlook}
\label{sec:conclusions}

The objective of this paper is to report on a recent analysis of the validity of standard linearized gravity, as it follows from GR. On the theoretical side, it is based on modifications of the EH action that are of mass dimension 6, involve the Riemann curvature tensor, and break diffeomorphism invariance through the presence of nondynamical background fields $k_R$ and $k_D$. This modified-gravity theory was linearized around the Minkowski metric to arrive at an altered wave equation for the metric perturbation. At that juncture, we focused on the term depending on $k_R$ and obtained modified dispersion relations that are degenerate for a small number of coefficients but nondegenerate for the majority of coefficients. The latter property is interpreted as gravitational-wave birefringence. Purely isotropic modifications of the propagation velocity were also identified, but most sectors, particularly the birefringent ones, exhibit anisotropies.

From the phenomenological side, the analysis rested upon time-of-flight data from both gravitational waves and photons. Nonbirefringent modifications of linearized gravity are excluded at the millimeter level, whereas the sensitivity to birefringence ranges down to 10 microns. This sensitivity is quite impressive when taking into account the experimental challenges of gravitational-wave detection. After all, detection has to be precise enough to measure the tiny strain amplitudes of a signal. However, the sensitivity for birefringence does not have to rely on the ability to measure spacetime strain, but rather on a sufficiently good time resolution of the measurement.

The constraints on nonbirefringent modifications are competitive since they are based on comparing gravitational-wave to photon data. Until now, only a single such event, GW170817/GRB 170817A, has been observed. Improvements are expected to be possible once there is a catalog of multiple similar events. Moreover, a better understanding of the mechanisms of intrinsic time delays between gravitational-wave and photon emission at the source, as well as the ability to quantify them depending on source properties, will be highly valuable. The birefringent bounds have the potential for improvement. A refined analysis of multiple events by means of linear programming \cite{Dantzig:1997,Diaz:2013wia,Kostelecky:2015dpa} or a waveform analysis based on Bayesian statistics can contribute to obtaining conservative limits on spacetime symmetry violation induced by the dim-6 coefficients $k_R$.

Recall also the background field $k_D$ of the second dim-6 operator. We have already determined the modified wave equation for the latter. The reasonable next step would be to derive the dispersion relations needed for a phenomenological survey. An alternative project could be to look into the $k_R^{(8)}RRR$ term; see Tab.~VI of Ref.~\cite{Kostelecky:2020hbb}. To do so, the systematic treatment of direct products of the Riemann tensor, presented in Ref.~\cite{Chung:2022ees}, could be indispensable.
These endeavors are likely to be more challenging than our present analysis, and they pose potential next steps in this scientific pursuit.

\section{Acknowledgments}

It is a pleasure to thank V.A.~Kosteleck\'{y} for valuable comments on certain aspects of the analysis. CMR acknowledges support from project Fondecyt Regular No.~1241369. CR is grateful for support from the Universidad San Sebastián 2026 Postdoctoral Researchers Attraction Program, grant USS-FIN-26-PDOC-03. MS is indebted to CNPq Produtividade 307653/2025-0 and CAPES/Finance Code 001.


\appendix

\section{Isotropic contributions}
\label{app:isotropic-configurations}

Here, we provide details on how to find the isotropic configurations of Eq.~\eqref{eq:isotropic-configurations}. To do so, we define the following $\mathit{SO}(3)$ tensors:
\begin{subequations}
\begin{align}
T_1^{ijkl}&:=\frac{1}{3}\delta^{ij}\delta^{kl}\,, \displaybreak[0]\\[2ex]
T_2^{ijkl}&:=\frac{1}{2}\left(\delta^{ik}\delta^{jl}+\delta^{il}\delta^{jk}-\frac{2}{3}\delta^{ij}\delta^{kl}\right)\,, \displaybreak[0]\\[2ex]
T_3^{ijkl}&:=\frac{1}{2}(\delta^{ik}\delta^{jl}-\delta^{il}\delta^{jk})\,.
\end{align}
\end{subequations}
The latter are orthogonal projectors, i.e.,
\begin{subequations}
\begin{align}
T_A^{ijkl}T_A^{klmn}&=T_A^{ijmn}\,, \\[2ex]
T_A^{ijkl}T_B^{klmn}&=0\,,
\end{align}
\end{subequations}
for $A,B\in\{1,2,3\}$. These projectors are useful for decomposing the sets of coefficients into traces and traceless parts.

\subsection{EE sector}

Compatibility with the symmetries implies the \textit{ansatz}
\begin{subequations}
\begin{equation}
K_{\text{EE}}^{\mathrm{ansatz}}=(\ring{k}_R)_1T_1+(\ring{k}_R)_2T_2+K_{EE}^{\mathrm{traceless}}\,,
\end{equation}
expressed as a function of the isotropic coefficients $(\ring{k}_R)_{1,2}$ governing each of the two isotropic sectors according to the group-theory finding of Eq.~\eqref{eq:irreps-EE}. Then,
\begin{align}
T_1\cdot K_{\text{EE}}^{\mathrm{ansatz}}&=(\ring{k}_R)_1=\frac{1}{3}\sum_{i,j} (k_R)^{0i0i0j0j}\,, \\[1ex]
T_2\cdot K_{\text{EE}}^{\mathrm{ansatz}}&=5(\ring{k}_R)_2 \notag \\
&=\sum_{i,j}\left((k_R)^{0i0j0i0j}-\frac{1}{3}(k_R)^{0i0i0j0j}\right)\,,
\end{align}
\end{subequations}
which implies $(\ring{k}_R)_{1,2}$ in terms of the controlling coefficients. Spatial indices have been suppressed for simplicity. A dot `$\cdot$' indicates a total contraction. Solving $K_{\text{EE}}=K_{\text{EE}}^{\mathrm{ansatz}}$ with $K_{\text{EE}}^{\mathrm{tracless}}=0$ implies the configurations of coefficients in terms of $(\ring{k}_R)_{1,2}$ that are associated with isotropic dispersion relations.
The nonzero SME coefficients of the EE sector are then chosen as
\begin{subequations}
\begin{align}
K_{\text{EE}}^{ii;ii}&=\frac{3}{5}(\ring{k}_R)_1\,,\quad K_{\text{EE}}^{ii;jj}=\frac{(\ring{k}_R)_2}{5}\,, \\[2ex]
(\ring{k}_R)_2&=\frac{2}{5}(\ring{k}_R)_1\,,
\end{align}
\end{subequations}
where $i\neq j$. Note that sums over equal spatial indices are not performed here. The explicit computation is best done with computer algebra. It leads to
\begin{equation}
\ring{\omega}_{\mathrm{EE}}=|\mathbf{p}|\left[1-\frac{16}{5}(\ring{k}_R)_1\mathbf{p}^2\right]\,,
\end{equation}
shared by both isotropic sectors. Thus, the first and second isotropic sectors mix with each other to provide an isotropic dispersion relation, and they cannot be separated.

\subsection{BB sector}

The BB sector is dual to EE and vice versa. Hence, the results to be found for BB resemble those of EE. We propose the \textit{ansatz}
\begin{subequations}
\begin{equation}
K_{\text{BB}}^{\mathrm{ansatz}}=(\ring{k}_R)_3T_1+(\ring{k}_R)_4T_2+K_{\text{BB}}^{\mathrm{traceless}}\,,
\end{equation}
such that
\begin{align}
T_1\cdot K_{\text{BB}}^{\mathrm{ansatz}}&=(\ring{k}_R)_3=\frac{1}{12}\sum_{i,j,k,l}(k_R)^{ijijklkl}\,, \\[1ex]
T_2\cdot K_{\text{BB}}^{\mathrm{ansatz}}&=5(\ring{k}_R)_4 \notag \\
&=\frac{1}{4} \sum_{i,j,k,l}\left((k_R)^{ijklijkl}-\frac{1}{3}(k_R)^{ijijklkl}\right)\,.
\end{align}
\end{subequations}
Solving $K_{\text{BB}}=K_{\text{BB}}^{\mathrm{ansatz}}$ with the traceless piece discarded provides an isotropic configuration of coefficients:
\begin{subequations}
\begin{align}
K_{\text{BB}}^{ii;ii}&=\frac{3}{5}(\ring{k}_R)_3\,,\quad K_{\text{BB}}^{ii;jj}=\frac{(\ring{k}_R)_4}{5}\,, \\[2ex]
(\ring{k}_R)_4&=\frac{2}{5}(\ring{k}_R)_3\,,
\end{align}
\end{subequations}
with $i\neq j$. Then, the isotropic dispersion relation is
\begin{equation}
\ring{\omega}_{\mathrm{BB}}=|\mathbf{p}|\left[1-\frac{16}{5}(\ring{k}_R)_3\mathbf{p}^2\right]\,.
\end{equation}
Similarly to the EE sector, the two isotropic BB sectors mix with each other and lead to a single isotropic dispersion relation.

\subsection{EB sector}

We make the \textit{ansatz}
\begin{subequations}
\begin{equation}
K_{\text{EB}}^{\mathrm{ansatz}}=(\ring{k}_R)_5T_1+(\ring{k}_R)_6T_2+K_{\text{EB}}^{\mathrm{traceless}}\,,
\end{equation}
such that
\begin{align}
T_1\cdot K_{\text{EB}}^{\mathrm{ansatz}}&=(\ring{k}_R)_5=\frac{1}{6}\sum_{i,j,k} (k_R)^{0i0ijkjk}\,, \\[1ex]
T_2\cdot K_{\text{EB}}^{\mathrm{ansatz}}&=5(\ring{k}_R)_6 \notag \\
&=\frac{1}{3}\sum_{i,j,k}(k_R)^{0i0ijkjk}-\sum_{i,j,k} (k_R)^{0i0jkikj}\,.
\end{align}
\end{subequations}
As before, the isotropic configurations are solutions of $K_{\text{EB}}=K_{\text{EB}}^{\mathrm{ansatz}}$, where the traceless part is set to zero:
\begin{subequations}
\begin{align}
K_{\text{EB}}^{ii;ii}&=\frac{(\ring{k}_R)_5}{5}\,,\quad K_{\text{EB}}^{ii;jj}=\frac{2}{5}(\ring{k}_R)_5\,, \displaybreak[0]\\[2ex]
K_{\text{EB}}^{ij;ij}&=\frac{(\ring{k}_R)_5}{10}\,,\quad (\ring{k}_R)_6=-\frac{(\ring{k}_R)_5}{5}\,,
\end{align}
\end{subequations}
for $i\neq j$. The dispersion relation then amounts to
\begin{equation}
\ring{\omega}_{\mathrm{EB}}=|\mathbf{p}|\left[1-\frac{64}{5}(\ring{k}_R)_5\mathbf{p}^2\right]\,.
\end{equation}

\subsection{EG and BG sectors}

Here, the following \textit{ans\"{a}tze} are proposed:
\begin{subequations}
\begin{align}
K_{\text{EG}}^{\mathrm{ansatz}}&=(\ring{k}_R)_7T_2+K_{\text{EG}}^{\mathrm{traceless}}\,, \displaybreak[0]\\[1ex]
K_{\text{BG}}^{\mathrm{ansatz}}&=(\ring{k}_R)_8T_2+K_{\text{BG}}^{\mathrm{traceless}}\,.
\end{align}
\end{subequations}
Applying the second projector leads to
\begin{subequations}
\begin{align}
T_2\cdot K_{\text{EG}}^{\mathrm{ansatz}}&=5(\ring{k}_R)_7=\frac{1}{2}\sum_{\substack{i,p \\ q,r}} \epsilon^{pqr}(k_R)^{0i0p0iqr}\,, \displaybreak[0]\\[1ex]
T_2\cdot K_{\text{BG}}^{\mathrm{ansatz}}&=5(\ring{k}_R)_8=\frac{1}{4}\sum_{\substack{i,j,k \\ m,n}} \epsilon^{kmn}(k_R)^{mnij0kij}\,.
\end{align}
\end{subequations}
In fact, nontrivial solutions to $K_{\text{EG}}=K_{\text{EG}}^{\mathrm{ansatz}}$ and $K_{\text{BG}}=K_{\text{BG}}^{\mathrm{ansatz}}$ were not found to exist. Thus, we conclude that both sectors are devoid of isotropic dispersion relations.


\end{document}